\documentclass[12pt,a4paper]{article}
\usepackage{a4wide}
\usepackage{latexsym}
\usepackage{subfigure}
\usepackage{epsf}
\usepackage{amssymb}
\makeatletter
\@addtoreset{equation}{section}
\makeatother

\begin{document}

\begin{flushright}
\small
WIS/27/04-DEC-DPP\\
{\bf hep-th/0412118}\\
December 12, 2004
\normalsize
\end{flushright}

\begin{center}


\vspace{.7cm}

{\Large {\bf Stability and Critical Phenomena of 

Black Holes and Black Rings}}

\vspace{1cm}


{\bf Giovanni Arcioni}
\footnote{E-mail: {\tt arcionig@phys.huji.ac.il }}\\[.2cm]

{\it Racah Institute of Physics,  
The Hebrew University of Jerusalem \\ 
Jerusalem 91904, 
Israel} \\[.7cm]

{\bf Ernesto Lozano-Tellechea}
\footnote{E-mail: {\tt ernesto.lozano@weizmann.ac.il}}\\[.2cm]

{\it Department of Particle Physics, 
Weizmann Institute of Science \\
Rehovot 76100, 
Israel}

\vspace{1cm}

{\bf Abstract}

\end{center}

\begin{quotation}

\small

We revisit the general topic of thermodynamical stability and critical
phenomena in black hole physics, analyzing in detail the phase diagram
of the five dimensional rotating black hole and the black rings
discovered by Emparan and Reall. First we address the issue of
microcanonical stability of these spacetimes and its relation to
thermodynamics by using the so-called Poincar\'e (or ``turning
point'') method, which we review in detail. We are able to prove that
one of the black ring branches is always locally unstable, showing
that there is a change of stability at the point where the two black
ring branches meet. Next we study divergence of fluctuations, the
geometry of the thermodynamic state space (Ruppeiner geometry) and
compute the appropriate critical exponents and verify the scaling laws
familiar from RG theory in statistical mechanics.  We find that, at
extremality, the behaviour of the system is formally very similar to a
second order phase transition.

\end{quotation}

\newpage
\tableofcontents

\pagestyle{plain}


\setcounter{footnote}{0}

\section{Introduction}

Topology change of event horizons constitutes a fundamental issue
in gravity. From a broad perspective, one of its most important
implications comes from its relation to the problem of resolution
of singularities. Instabilities leading to the destruction of an
event horizon can in principle produce naked singularities, hence
violating the Cosmic Censorship Conjecture. The latter constitutes
one of the major unsolved issues of the classical theory, and a
complete understanding of this problem must imply deep
consequences at the quantum level. Somehow, the Cosmic Censorship
can be thought as the classical mechanism for resolution of
singularities, which ultimately are expected to be cured by any
correct quantum completion of gravity.

In a String Theory context, singularity resolution involving a
(manifold) topology change was first emphasized
in~\cite{Strominger:1995cz}, where the inclusion of nonperturbative
string states (namely, black hole states) was shown to provide a
mechanism for the resolution of the conifold
singularity. Gravitational phase transitions also play an important
role in the $AdS$/CFT correspondence, starting from the original work
of Witten~\cite{Witten:1998zw} on nucleation of black holes in
Anti-de-Sitter space (the Hawking-Page phase
transition~\cite{Hawking:1982dh}) and its relation to deconfinement at
finite temperature in gauge theories. This is of interest both for the
study of the dual phenomena in the corresponding gauge theory (recent
works include e.g.~\cite{Aharony:2003sx}) and, also, because it seems
to provide the possibility of investigating gravitational
singularities from the field theory point of view~\cite{Kraus:2002iv}.
\newline

Interestingly, higher dimensional General Relativity provides
indications of topology chan\-ge of event horizons, the best known
example being the Gregory-Laflamme
instabili\-ty~\cite{Gregory:vy}. Recently this phenomenon has been
studied in the presence of compact dimensions by a number of
authors~\cite{Harmark:2003eg} (for a recent review
see~\cite{Kol:2004ww}). In general, the question of topology change of
event horizons makes sense whenever different phases of black objects
(black holes, black strings, black $p$-branes or others) can exist, a
general feature of higher dimensional General Relativity and
low-energy String Theory.

Of course, and at least at classical level, the key issue to be
investigated about possible transitions between different black phases
concerns their {\em dynamical} stability, i.e.~stability of the
solution under field perturbations, or classical stability under
Lorentzian time evolution. However, not much is known about the
classical stability of solutions of GR above four dimensions. Apart
from the well known instabilities of the Gregory-Laflamme type, very
little is known about many other solutions of interest, like
e.g. those with angular momentum. For those, some partial or indirect
checks of dynamical stability have been put
forward~\cite{Emparan:2003sy} but, in most cases, an exhaustive and
rigorous proof of (in)stability is still lacking.

A way to address the problem of stability of event horizons comes from
black hole thermodynamics. In ordinary thermodynamics of extensive
systems, local {\em thermodynamical} stability (defined as equivalent
to the condition that the Hessian of the entropy has no positive
eigenvalues) is related to the {\em dynamical} stability of the
system. However, such a parallel cannot be inferred for black holes,
which constitute non-extensive, non-additive thermodynamical
systems. The reasons for this depart from standard thermodynamics can
be phrased in different ways: the long-range behaviour of
gravitational forces, the well known scaling of the BH entropy with
the area (rather than the volume), or the fact that BHs constitute
elementary entities that cannot be subdivided into separate systems,
not even in any idealized manner.

Critical thermodynamic behaviour of black holes was first studied
in~\cite{Davies:1978mf} and, since then, by a number of authors (see
e.g.~\cite{Lousto:1996kg} and references therein). However, the
interpretation of the divergences in the specific heats and
generalized susceptibilities (in particular, their interpretation as a
true phase transition and their relation to the stability of the
system) is not clear and has been subject of
debate~\cite{Curir,Sorkin:1982ut,Pavon:1988in,Kaburaki:TSKerrBH,
Katz:1993up,KabuKerr,kabuscaling}.  In a String Theory context some
partial attempts to relate dynamical stability and local
thermodynamical stability have been put forward as
well~\cite{Gubser:2000mm}, but currently the present status of this
subject is not clear~\cite{Marolf:2004fy} (see also the third
reference listed in~\cite{Aharony:2003sx}). We will comment on the
relation between the Gubser-Mitra conjecture and the ``Poincar\'e
method'' that we will use in this paper in the Conclusions.\newline

In this paper we will explore these issues in the context of five
dimensional gravity. This scenario seems interesting to us due to the
discovery made by Emparan and Reall of a new BH phase: an
asymptotically flat, rotating black hole solution with horizon
topology $S^1\times S^2$ and carrying angular momentum along the $S^1$
--- what they called a ``black ring''~\cite{Emparan:2001wn}. Such
black ring solutions exist in a region in parameter space (i.e. mass
and angular momentum) which overlaps with that of the five dimensional
spherical rotating black hole with rotation in just one plane studied
in~\cite{Myers:un}. This fact provides the first known example of BH
non-uniqueness in asymptotycally flat space, and it raises questions
on stability and on the possibility of transitions between the
different phases. 

Let us summarize the main properties of these spacetimes. The five
dimensional rota\-ting black hole~\cite{Myers:un} has an upper ``Kerr
bound'' in its angular momentum per unit mass, the extremal
configuration that saturates the bound being a naked singularity. In
what follows, we will be using the dimensionless quantity
\begin{displaymath}
  x^2 \sim \frac{J^2}{M^3}
\end{displaymath}
(where $J$ and $M$ are respectively the total angular momentum and
mass) as the ``control parameter''\footnote{A proper {\em order}
parameter will be defined in Section~\ref{sec:ScalingLawsBHBR}.} of
the problem --- analogous, say, to the temperature in a liquid-gas
system when studied in the canonical or grand canonical ensembles. The
same parameter has been already used in investigations of black rings
in~\cite{Emparan:2004wy}. It is normalized in such a way that the five
dimensional rotating black hole solution exists in the range
\begin{displaymath}
  {\rm Black \ Hole:} \ \ \ \ \ 0\leq x<1\, ,
\end{displaymath}
$x=1$ being the extremal (singular) limit. On the other hand, one can
have {\em two} different black ring spacetimes with horizon topology
$S^1\times S^2$ when the total angular momentum per unit mass exceeds
a certain minimal value $x_{\rm min}$ given by:
\begin{displaymath}
  x_{\rm min}=\sqrt{\frac{27}{32}}\approx 0.919\, .
\end{displaymath}
One of them, that we will call the ``large'' black ring, can have
angular momentum per unit mass unbounded from above. The other one,
that we call ``small'' black ring, cannot\footnote{The labels
``large'' and ``small'' for the two different black rings have already
been introduced in~\cite{Emparan:2004wy}.}. One has:
\begin{displaymath}
  \left\{
  \begin{array}{rl}
    {\rm Large \ Black \ Ring:} \ \ \ \ \ &
    x_{\rm min}\leq x<\infty \, , \\[.3cm]
    {\rm Small \ Black \ Ring:} \ \ \ \ \ &
    x_{\rm min}\leq x< 1\, .
  \end{array}
  \right.
\end{displaymath}
At $x_{\rm min}$ both solutions coincide. For a comparison of the
respective entropies see Fig.~\ref{fig:Entropies}. One can see that
near extremality the large black ring is entropically favoured (its
entropy equals that of the rotating BH at $x = 2\sqrt{2}/3 \approx
0.943$). This fact was taken in Ref.~\cite{Emparan:2001wn} as a first
indication of a phase transition from the black hole to a black ring
when the former rotates fast enough, i.e. as we vary $x$. Even if such
a topology changing process is not believed to be possible
classically~\cite{Horowitz:2001cz}, one might expect to find some
evidence of it by exploring issues like stability and the
thermodynamics of these spacetimes. The study of the general
properties of this phase diagram is the main purpose of this
paper. \newline

There are two main concrete questions that we will address here. The
first one is that of the classical stability of these solutions and
its relation to thermodynamics. We will deal with this issue in
Section~\ref{sec:DvsThStab}, and our main tool will be the so-called
``Poincar\'e'' or ``turning point'' method.

The second question is that of to which extent we are allowed to call
all these configurations ``phases'' in any thermodynamical or
statistical mechanical sense. This is a nontrivial question which has
to do with the properties of the classical ``configuration space''
(i.e. the space of metrics): namely, if all these solutions correspond
or not to extrema of some common underlying thermodynamic
potential. If so, the general theory of critical phenomena tells us
that, near a critical point, the generalized susceptibilities and
fluctuations have to diverge in a very specific way.  We will address
these issues in Sections~\ref{sec:FCBAndTG}
and~\ref{sec:ScalingLawsBHBR}. Our main tools will be the study of the
geometry of the thermodynamic space state and the general theory of
second order phase transitions.

\subsubsection*{Summary of Results}

After reviewing in Section~\ref{sec:RevBHAndBR} the main properties of
rotating black holes and black rings, we focus in
Section~\ref{sec:DvsThStab} on the issue of stability. We revise the
usual arguments relating dynamical and local thermodynamical
stability, and recall why these cannot be applied to non-extensive
systems such as BHs. Instead, we will use the so-called ``Poincar\'e
method of stability''~\cite{poincare} as it has been already applied
to BH physics
in~\cite{Kaburaki:TSKerrBH,Katz:1993up,KabuKerr,Concave}. This
technique is of remarkable simple applicability, and it essentially
consists on plotting the appropriate phase diagrams. We will review it
in detail and comment on the suitability of this method for the study
of BHs. Contrary to the standard analysis based on the sign of the
specific heat, the Poincar\'e method does {\em not} predict any
instability in e.g. the Schwarzschild or Kerr spacetimes. We will use
it to study the BH/BR system in the microcanonical ensemble, finding
that the only point at which a change of stability can occur is at
$x_{\rm min}$. One of our main results is in fact that the small black
ring is not only globally, but also {\em locally} unstable.  The large
black ring turns out to be a ``more stable'' phase in a sense that we
will explain. On the other hand, we do not find any change of stability
along the whole BH branch. In particular, we will see that nothing
special happens at the point in the BH branch where the specific heat
changes sign (that is, the five dimensional analogue of the ``critical
point'' found by Davies in~\cite{Davies:1978mf} and discussed since
then by many other authors). 

In Section~\ref{sec:FCBAndTG} we discuss divergence of fluctuations
and critical behaviour. We will see that quadratic fluctuations can be
evaluated at $x_{\rm min}$ and $x=1$, due to the fact that at those
points the relevant degrees of freedom of the system are just a few
and under control. We will see that at both points fluctuations
diverge, but that the origin of this divergent behaviour is very
different at $x_{\rm min}$ and at extremality. Next we will use the
geometry of the thermodynamic state space as introduced by
Ruppeiner~\cite{Rupp}, and explain how this technique should be
interpreted and applied to the BH case. In particular, the divergences
of the ``thermodynamic curvature'' are interpreted in Ruppeiner theory
as indicators of critical phenomena. We will compute this quantity for
the BH/BR system, finding that it diverges both at $x_{\rm min}$ and
$x=1$.

In Section~\ref{sec:ScalingLawsBHBR} we compute the various critical
exponents at the points where fluctuations diverge. We check the usual
scaling laws familiar from Renormalization Group theory in statistical
mechanics, finding that these are obeyed at extremality, but not at
$x_{\rm min}$. We thus conclude that $x=1$ might be identified, in a
certain sense that we will explain, with the critical point of a
second order phase transition. The phenomenon at $x_{\rm min}$ seems,
on the other hand, very different: there fluctuations diverge because,
as a consequence of the change of stability, the thermodynamic
potential becomes locally flat. We discuss our results and prospects
for future work in Section~\ref{sec:Conclusions}.


\section{The Black Hole and Black Ring Spacetimes}
\label{sec:RevBHAndBR}

\subsection{The $d=5$ Rotating Black Hole}

In~\cite{Myers:un} Myers and Perry found black hole solutions of
higher dimensional GR with arbitrary angular momenta. In particular,
the $d=5$ BH solution with rotation in a single plane is found to be
given by the line element:
\begin{equation}
  \label{MPBH}
  ds^2 = -dt^2 + \Delta(dt-a\sin^2\theta\, d\psi)^2
   + \sin^2\theta\, (r^2+a^2)d\psi^2 + \Psi\, dr^2
   + \rho^2\, d\theta^2 + r^2\cos^2\theta d\phi^2\, ,
\end{equation}
with $0\leq\psi\leq 2\pi$, $0\leq\phi\leq 2\pi$,
$0\leq\theta\leq\pi/2$ and the functions $\Delta$, $\Psi$ and $\rho$
being:
\begin{equation}
  \label{DeltaPsirho}
  \Delta \equiv \displaystyle \frac{\mu}{r^2+a^2\cos^2\theta}\, , \ \
  \ \ \Psi \equiv \displaystyle
  \frac{r^2+a^2\cos^2\theta}{r^2+a^2-\mu}\, , \ \ \ \ \rho^2 \equiv
  r^2 + a^2\cos^2\theta\, .
\end{equation}
The parameters $\mu$ and $a$ set the physical ADM mass and angular
momentum via the relations:
\begin{equation}
  \label{PhysicalMAndJ}
  M = \frac{3\pi}{8G}\, \mu\, , \ \ \ \ \ \ \ \ \
  J = \frac{2}{3}Ma\, .
\end{equation}
This solution has a ``ring singularity'' at $\rho=0$ and a event
horizon (of spherical topology) at $r_H^2 = \mu-a^2$, whose existence
thus requires the bound
\begin{equation}
  \label{ExtremalityBound}
  a^2<\mu \ \ \Longleftrightarrow \ \ \frac{J^2}{M^3}<\frac{32G}{27\pi}\, .
\end{equation}
The extremal case ($\mu=a^2$) is a naked singularity. 

\subsubsection*{Thermodynamic Quantities}

Let us introduce the following dimensionless quantity (a suitably
normalized angular momentum ``per unit mass''):
\begin{equation}
  \label{x}
  x \equiv \left(\frac{27\pi}{32G}\right)^{1/2}\frac{J}{M^{3/2}}
    = \frac{a}{\sqrt{\mu}}\, .
\end{equation}
In terms of it and the mass parameter $\mu$, the area, temperature
and angular velocity of the horizon are given by:
\begin{equation}
  \label{BHEntropy}
  {\cal A}_{\rm BH} = 2\pi^2\, \mu^{3/2}\, \sqrt{1-x^2}\, ,
  \ \ \ \ \ \ \ \
  T_{\rm BH} = \frac{1}{2\pi\, \sqrt{\mu}}\, \sqrt{1-x^2}\, ,
  \ \ \ \ \ \ \ \
  \Omega_{\rm BH} = \frac{x}{\sqrt{\mu}}\, .
\end{equation}
In particular, one can see that the area decreases when
$a^2\rightarrow\mu$ at fixed mass (i.e. $x\rightarrow 1$), and it
vanishes precisely in the extremal limit. This is because, as we
increase the angular momentum of the BH, the event horizon gets
``flattened'', becoming a (singular) disc of zero area in the extremal
limit (for an explicit analysis of this geometry see
e.g.~\cite{Emparan:2003sy}).

\subsection{The Black Ring Spacetime}
\label{sec:TheBRSpacetime}

Let us review now the solution found by Emparan and Reall
in~\cite{Emparan:2001wn}\footnote{Note however that here we will be
using the conventions of~\cite{Elvang:2003mj}.}. This solution
describes a five dimensional, asymptotically flat black hole whose
horizon has $S^1\times S^2$ topology. The solution is regular on and
outside the event horizon provided that it has angular momentum along
the $S^1$ direction. Both the black ring and the rotating black hole
solution are described by the metric:
\begin{equation}
  \label{BRxy}
  \begin{array}{l}
  ds^2 = \displaystyle
         -\frac{F(x)}{F(y)}
         \left(dt + R\sqrt{\lambda\nu}\, (1+y)\, d\psi\right)^2 + \\[.5cm]
  \displaystyle \ \ \ \ \ \ \ \ \,
         +\frac{R^2}{(x-y)^2}\left[
         -F(x)\left(G(y)\, d\psi^2 +
         \frac{F(y)}{G(y)}\, dy^2\right)
         +F^2(y)\left(\frac{dx^2}{G(x)}+\frac{G(x)}{F(x)}\, d\phi^2\right)
         \right]\, ,
  \end{array}
\end{equation}
where the functions introduced above are:
\begin{equation}
  F(\xi)=1-\lambda\xi\, , \ \ \ \ \ \ \ \
  G(\xi)=(1-\xi^2)(1-\nu\xi)\, .
\end{equation}
$R$ and $\nu$ are parameters whose appropriate combinations give the
total ADM mass and angular momentum, while $\lambda$ has to take
specific values in order to avoid conical singularities. In
particular, there are two possibilities:
\begin{equation}
  \label{lambdaBHlambdaBR}
  \left\{
  \begin{array}{rcll}
  \lambda&=&1\, ,  & {\rm (black\ \ hole)}\, , \\[.2cm]
  \lambda&=& \displaystyle
            \frac{2\nu}{1+\nu^2}\, , & {\rm (black\ \ ring)}\, .
  \end{array}
  \right.
\end{equation}
Once $\lambda$ is fixed one has therefore two parameters: $R$ (with
dimensions of length and setting the ``radius'' of the BH or the black
ring) and $\nu$ (dimensionless and varying from 0 to 1).  In the BH
case $\nu=0$ represents a non rotating black hole, while $\nu=1$
yields the extremal case. In the BR case, and as explicitly shown
below, as $\nu$ takes values from 0 to 1 we get {\em two} different
black ring configurations, which coincide only at $\nu=1/2$. At
$\nu=0$ we have an infinitely thin and large black ring with infinite
angular momentum per unit mass. The angular momentum per unit mass of
this object takes its minimal value at $\nu=1/2$. From $\nu=1/2$ to
$\nu=1$ the solution describes a second black ring of different
geometry, which approaches the singular BH spacetime in the extremal
limit $\nu\rightarrow 1$. At $\nu=1/2$ both solutions coincide,
yielding the black ring spacetime with minimum possible angular
momentum per unit mass. These properties can be visualized e.g. from
the entropy diagram of Fig.~\ref{fig:Entropies}. 

In~\cite{Elvang:2003mj} the physical parameters of both the black ring
and black hole solutions are found to be:
\begin{equation}
  \label{PhysicalParameters}
  \left\{
  \begin{array}{rclcrcl}
    {\cal A}&=&8\displaystyle
         \pi^2 R^3 \frac{\sqrt{\lambda}(1+\lambda)(\lambda-\nu)^{3/2}}
                      {(1+\nu)^2(1-\nu)}\, , &\ \ \ \ &
    M&=&\displaystyle
        \frac{3\pi R^2}{4G}\frac{\lambda(\lambda+1)}{\nu+1}\, , \\[.4cm]
    T&=&\displaystyle \frac{1}{4\pi R}\frac{1-\nu}
                              {\sqrt{\lambda(\lambda-\nu)}}\, , &\ \ \ \ &
    J&=&\displaystyle
        \frac{\pi R^3}{2G}\frac{\sqrt{\lambda\nu}(1+\lambda)^{5/2}}
                               {(1+\nu)^2}\, . \\[.4cm]
    \Omega&=&\displaystyle
             \frac{1}{R}\sqrt{\frac{\nu}{\lambda(1+\lambda)}}\, .
  \end{array}
  \right.
\end{equation}
We recover the black hole or black ring values by
using~(\ref{lambdaBHlambdaBR}). Finally, let us recall that the
physical parameters in~(\ref{PhysicalParameters}) satisfy the Smarr
relation given by
\begin{equation}
  \label{Smarr}
  M = \frac{3}{2}\left(\frac{T{\cal A}}{4G}+\Omega J\right)\,
\end{equation}
for any value of $\lambda$, which in particular means that both the
black hole and the black ring spacetimes satisfy the same Smarr
relation.

\subsubsection*{Thermodynamic Quantities}

In order to compare entropies and eventually carry out a thermodynamic
analysis, we will need a fundamental relation ${\cal A}={\cal A}(M,J)$
and the corresponding ``equations of state'' for the black ring
spacetime too. Defining the parameters $\mu$ and $x$ in terms of the
physical mass and angular momentum as in~(\ref{PhysicalMAndJ})
and~(\ref{x}), one finds:
\begin{equation}
  \label{BRFundamentalRelation}
  {\cal A}_{\rm BR} = \sqrt{2}\pi^2\, \mu^{3/2}\,
                      \sqrt{\nu(1-\nu)}\, ,  \ \ \ \
  T_{\rm BR} = \frac{1}{2\sqrt{2}\pi\, \sqrt{\mu}}\,
               \sqrt{\frac{1-\nu}{\nu}}\, ,  \ \ \ \
  \Omega_{\rm BR} = \frac{\sqrt{2}}{\sqrt{\mu}}\,
                    \sqrt{\frac{\nu}{1+\nu}}\, ,
\end{equation}
where $\nu=\nu(J^2/M^3)$ is given by any of the following two values:
\begin{equation}
  \label{nuLS}
  \left\{
  \begin{array}{rcl}
  \nu_S&=&\sqrt{8}\, x\, \left(\frac{1}{\sqrt{3}}\, \cos\Theta
          + \sin\Theta \right) - 1\, , \\
  \nu_L&=&\sqrt{8}\, x\, \left(\frac{1}{\sqrt{3}}\, \cos\Theta
          - \sin\Theta \right) - 1\, ,
  \end{array}
  \right.
\end{equation}
the angle $\Theta$ being
\begin{equation}
  \label{Theta}
  \Theta=\frac{1}{3}\arctan\sqrt{\frac{32}{27}\, x^2-1}\, .
\end{equation}
These explicit expressions are obtained by inverting the
parametrization from~(\ref{PhysicalParameters}) and taking into
account the conditions to avoid naked singularities\footnote{This
fundamental relation has already been given in parametric form
in~\cite{Emparan:2004wy}.}. The fact that the angular momentum at
fixed mass is bounded from below at at the precise value $x_{\rm
min}=\sqrt{27/32} \approx 0.919$~\cite{Emparan:2001wn} is explicit in
the formula above.

In~(\ref{nuLS}) $\nu_S$ and $\nu_L$ correspond to the two possible
black rings found in~\cite{Emparan:2001wn}: $\nu_S$ corresponds to
what we will call ``small'' ring (i.e. the one whose angular momentum
is also bounded from above at $x=1$), while $\nu_L$ corresponds to the
``large'' black ring (the one with unbounded angular momentum).  As
explained, in the limit $x\rightarrow\infty$ the horizon of this large
black ring takes the shape of an infinitely large (along the $S^1$)
and thin ring, hence the name. The fact that the angular momentum of
the small BR is also bounded from above, while the one of the large BR
remains unbounded, can be deduced by inspection of~(\ref{nuLS}) and
recalling that $\nu$ is constrained to take values in
$[0,1)$~\cite{Elvang:2003mj}.

\subsection{Entropy Diagrams}

\begin{figure}[!ht]
\begin{center}
\leavevmode
\epsfxsize= 8cm
\epsffile{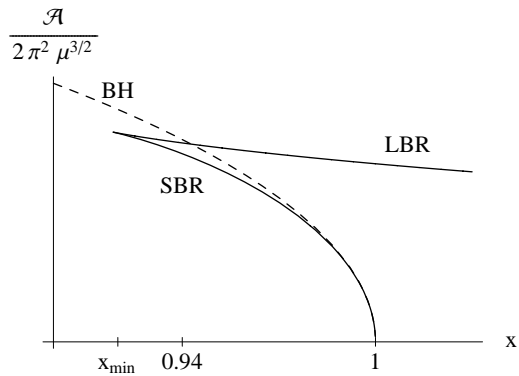}
\caption{\footnotesize{Plot of ${\cal A}/2\pi^2\mu^{3/2}$ against
$x=\sqrt{27\pi/32G}J/M^{3/2}$ for the Myers-Perry BH (dashed line) and
the two black rings (see~\cite{Emparan:2001wn}). At $x_{\rm min} =
\sqrt{27/32}\approx 0.919$ black ring formation becomes possible. At
$x=2\sqrt{2}/3\approx 0.943$ the entropy of the large BR exceeds 
that of the black hole.}}
\label{fig:Entropies}
\end{center}
\end{figure}

The expressions~(\ref{BHEntropy}) and~(\ref{BRFundamentalRelation})
allow for an explicit comparison of the respective entropies. We
recover the plot already found in~\cite{Emparan:2001wn}, which is
shown in Fig.~\ref{fig:Entropies}.

This figure is the first indication of the possibility of a phase
transition when we vary the angular momentum at fixed mass. As we see,
beyond $x=2\sqrt{2}/3\approx 0.942$ the entropy of the large black
ring exceeds the one of the black hole, which rapidly decreases to
zero at $x=1$. Emparan and Reall pointed out in their original work
that this fact may be taken as an indication of a BH/BR phase
transition. 

Motivated by this fact, we will study the stability properties and the
behaviour of fluctuations of these spacetimes. However, our results
will lead us to the conclusion that, from the point of view of
stability, the relevant points in this phase diagram are $x_{\rm min}$
and $x=1$, and that nothing special seems to be happening at
$x=2\sqrt{2}/3$. We will comment on the implications of this result in
the Conclusions.


\section{Dynamical vs. Thermodynamical Stability}
\label{sec:DvsThStab}

To argue in favour or against a possible phase transition, the first
issue we should address is that of the stability of the classical
solutions. To face this problem, one may think that a way to
circumvent the study of linear perturbations of the metric might come
from thermodynamics. In ordinary thermodynamics of extensive systems,
local {\em thermodynamical} stability (defined as the Hessian of the
entropy having no positive eigenvalues) is linked to the {\em
dynamical} stability of the system: a positive mode in the Hessian
means that at least some kind of small fluctuations within the system
are entropically favoured, implying that the system is unstable
against these. However, the simple argument leading to this
conclusion~\cite{Callen} relies on the {\em additivity} of the entropy
--- a property which does not hold for black holes. The best known
example of this failure is Schwarzschild black hole: a stable
configuration with negative specific heat (i.e. a positive
Hessian). \newline

In order to make clear the differences with the case of BH
thermodynamics, let us first review the standard criteria of local
thermodynamical stability. To fix ideas, consider an extensive
thermodynamic system in the microcanonical ensemble with a fundamental
relation $S=S(M,J)$ (where $M$ and $J$ stand for the internal energy
and some other extensive parameter). Divide it into two identical
subsystems and consider a spontaneous fluctuation that transfers some
amount of energy $dM$ from one subsystem into the other. Then, if the
system is to be stable, one requires the total entropy after this
energy transfer to be smaller (or equal) than the previous total
entropy:
\begin{equation}
  S(M+dM,J)+S(M-dM,J) \leq 2S(M,J)\, ,
\end{equation}
since, otherwise, energy transfers within the system will take place
spontaneously, therefore meaning that the system would be unstable
against redistributions of mass from homogeneity (this would be the
dynamical unstable mode). The differential form of the above condition
is:
\begin{equation}
  \label{Stab1}
  H_{MM} \equiv \frac{\partial^2S}{\partial M^2} \leq 0\, .
\end{equation}
So~(\ref{Stab1}) is a necessary condition for a system to be
stable. Similar considerations lead to
\begin{equation}
  \label{Stab2}
  H_{JJ} \equiv \frac{\partial^2S}{\partial J^2} \leq 0\,
\end{equation}
for transfers of $J$ within the system. For ``cooperative'' transfers
of both $M$ and $J$ we have to require the Hessian $H$ of $S(M,J)$ to
not to be positive definite. Hence
\begin{equation}
  \label{Stab3}
  \det{H} =
  \frac{\partial^2 S}{\partial M^2}\,
  \frac{\partial^2 S}{\partial J^2}
  - \left(\frac{\partial^2 S}{\partial M\partial J}\right)^2 \geq 0\, .
\end{equation}
So conditions~(\ref{Stab1})-(\ref{Stab3}) have to be satisfied if the
system is to be stable. Only two of them are necessary, the
``invariant'' condition being a restriction on the sign of the two
eigenvalues of the Hessian matrix (for example, (\ref{Stab1}) and
(\ref{Stab3}) imply (\ref{Stab2})).

In standard thermodynamics the violation of any of the above
inequalities is taken as an indication of instability and a phase
transition. They are equivalent to the more familiar criteria of
positivity of the various specific heats and susceptibilities (for
example, (\ref{Stab1}) implies positivity of the specific heat at
constant $J$, etc.)

\subsection{Stability of Black Holes and Non-Extensive Systems}
\label{sec:BHsAndNES}

Note that the above argument relies on the additivity property of the
entropy when merging subsystems and, more fundamentally, on the very
possibility of describing a given system as made up of constituent
subsystems. Standard thermodynamical stability criteria fail in
general when applied to black holes because their entropy is
non-additive (the BH entropy is never an homogeneous first-order
function of its variables), and because in GR a local definition of
mass and other familiar ``should-be-extensive'' quantities is not
possible. Therefore, BHs cannot be thought as made up of any
constituent subsystems each of them endowed with its own
thermodynamics, and as such they are non-extensive objects. Let us
recall that extensivity of the entropy is a postulate in ordinary
thermodynamics~\cite{Callen}, and many general results in
thermodynamic theory just fail if this property is not assumed.

The study of non-extensive thermodynamics has been worked out to some
extent, though (see e.g.~\cite{Landsberg} and references therein), and
a concrete proposal to avoid the problems just mentioned in the BH
case has been put forward in a series of articles by several
authors~\cite{Kaburaki:TSKerrBH,Katz:1993up,KabuKerr}. These works are
based on the so-called ``Poincar\'e method of
stability''~\cite{poincare}. This method was previously applied to
gravitating systems in~\cite{katzI}, where it was shown to work in the
task of showing up the existence of unstable modes of an equilibrium
solely from the properties of the equilibrium sequence, without having
to solve any eigenvalue equation. Formal proofs of the method include
those given by Katz~\cite{katzI} and
Sorkin~\cite{Sorkin:1981jc,Sorkin:1982ut}. A somewhat alternative and
more intuitive interpretation has been given in~\cite{Concave}. Let us
review these stability analyses before applying them to the BH/BR
system.

\subsubsection*{Setup}

As emphasized in~\cite{Concave}, one has to consider a single black
hole as a whole system.  In this way, any discussion related to the
possibility of dividing it into subsystems or to the additivity
property of the entropy simply does not take place. Second, one has to
put the black hole in a given thermodynamic ensemble. The values of
the fixed thermodynamic quantities that specify the ensemble will be
the usual {\em global} quantities that one can define in GR, thus
avoiding any problem related to the non-locality of their
definitions. This is just what we will be doing if, for example, the
chosen ensemble is the microcanonical and our starting point is the
fundamental relations~(\ref{BHEntropy}) and
(\ref{BRFundamentalRelation}).

Considering the BH as a whole system means that, for instance,
negative specific heat at constant $J$ only indicates an instability
of the canonical ensemble: small interchanges of energy between the BH
and an {\em external} thermal bath render the system
unstable. However, we do not expect such considerations to be related
to the issue of classical stability of the gravitational solution
under small perturbations.

\subsubsection*{Equilibrium States, Entropy Maximum Postulate and
Fundamental Relations}

The next step is to make a stability analysis based only on
maximization of entropy. In particular, we do not want to assume
additivity of the entropy function. Before proceeding to such an
analysis, let us first make some definitions and set some notations
(we follow closely Sorkin~\cite{Sorkin:1981jc,Sorkin:1982ut}).
\begin{itemize}
\item Let ${\cal M}$ be the space of possible states or configurations
of a given system. A given configuration is specified by a point $X$
in ${\cal M}$. Let $\{\mu^i\}$ denote the set of independent
thermodynamic variables that specify a given ensemble, and ${\cal S}$
be the corresponding Massieu function\footnote{In the microcanonical
ensemble the $\{\mu^i\}$ are variables like the energy and the Massieu
function is the entropy. In other ensembles, the Massieu function is
given by the Legendre transform of the entropy.}.  Both ${\cal S}$ and
the set of $\{\mu^i\}$ are functions on ${\cal M}$ which we assume to
be smooth (for our purposes, having at least continuous second
derivatives) except maybe in some boundary of ${\cal M}$\footnote{In
practice this will mean ``extremality'' (see below).}. 

\item {\em Equilibrium states} (stable or not) occur at points in
${\cal M}$ which are extrema of ${\cal S}$ under displacements $dX$
for which $d\mu^i=0$. At any equilibrium state one can define {\em
conjugate thermodynamic variables} $\{\beta_i\}$ such that, for all
displacements $dX$,
\begin{equation}
  \label{VariationMassieuFunction}
  d{\cal S} = \beta_id\mu^i\, , 
\end{equation}
reflecting the fact that, {\em at equilibrium}, the Massieu function
only depends on the values of the thermodynamic variables
$\{\mu^i\}$. Unless the relations $d\mu^i=0$ are not linearly
independent, Eq.~(\ref{VariationMassieuFunction}) defines the
quantities $\{\beta_i\}$ uniquely. The set of equilibria are thus a
submanifold ${\cal M}_{\rm eq}$ of the configuration space. Curves in
this submanifold are called ``equilibrium sequences'', and points in
${\cal M}_{\rm eq}$ can be labeled by the corresponding values of the
$\{\mu^i\}$, which are referred to as ``control
parameters''\footnote{These names come from catastrophe theory, which
is the mathematical framework for the study of the phenomena will be
interested in. See e.g.~\cite{Arnold}}. The points in ${\cal
M}\setminus{\cal M}_{\rm eq}$ are called {\em off-equilibrium}
states\footnote{More properly, one should speak about {\em
constrained} equilibria, since in principle an entropy function is
only guaran\-teed to exist at equilibrium. It is of course difficult
to say what ``constrained equilibria'' means in a gravitational
setting.  In this respect, we will have to assume that thermodynamic
functions are somehow well defined for off-equilibrium states (see
e.g.~\cite{HK77,Katz:2000vv}), at least in the vicinity of ${\cal
M}_{\rm eq}$.}.

\item The {\em entropy maximum postulate} states that unconstrained
locally {\em stable} equilibrium states take place at points in ${\cal
M_{\rm eq}}$ where ${\cal S}$ is a local maximum w.r.t. arbitrary
variations $dX$ in ${\cal M}$ that preserve $d\mu^i=0$.

\item A {\em fundamental relation} for the equilibrium states is an 
explicit expression for the function
\begin{equation}
  \label{GeneralFundamentalRelation}
  {\cal S}_{\rm eq}={\cal S}(\mu^i)\, ,
\end{equation}
i.e. the integral of~(\ref{VariationMassieuFunction}).  Given a
fundamental relation, the variation of the Massieu function as given
by Eq.~(\ref{VariationMassieuFunction}) serves to calculate explicit
expressions for the conjugate thermodynamic variables $\{\beta_i\}$
{\em at equilibrium}:
\begin{equation}
  \label{GeneralEquationsOfState} 
  \beta_i(\mu^j)=\frac{\partial{\cal S}_{\rm eq}}
                      {\partial\mu^i}\, .
\end{equation}
The above equations constitute the {\em equations of state} of the
system.
\end{itemize}

The entropy maximum postulate is therefore a statement about the
behaviour of ${\cal S}$ {\em along off-equilibrium curves} in ${\cal
M}$, not about its variations along ${\cal M}_{\rm eq}$. In general,
the properties of ${\cal S}$ along ${\cal M}_{\rm eq}$ do not tell us
anything about the local stability of an equilibrium, and expressions
like~(\ref{Stab1}) where $S$ is the equilibrium entropy do not provide
information on local stability if one cannot invoke additivity of the
entropy over constituent subsystems. However, the equilibrium entropy
as given by~(\ref{GeneralFundamentalRelation}) is often all what we
know. In particular, it all what we know when considering BHs, because
expressions such as~(\ref{BHEntropy}) or~(\ref{BRFundamentalRelation})
only provide values of the entropy function restricted to a very
particular region of configuration space: namely, the series of
(equilibrium) black hole solutions.

Therefore, questions about local stability based on maximization of
entropy (or Massieu function) are not questions about the properties
of the restricted function ${\cal S}_{\rm eq}$. What one needs is an
{\em extended Massieu function}, henceforth denoted by $\widehat{\cal
S}$, that reflects the behaviour of the quantity ${\cal S}$ off the
black hole states, at least in their vicinity.  However, the form of
$\widehat{\cal S}$ is not known in general. To have it, one would need
a parametric family of spacetimes describing deformations of the black
hole series, and then to compute all the relevant thermodynamic
quantities from them.

\subsubsection*{The ``Turning Point Method''}

The so-called ``Poincar\'e''~\cite{poincare} or ``turning point''
method allows, under certain assumptions, to infer some information
about the behaviour of $\widehat{\cal S}$ and on stability by using
only the {\em equilibrium} equations of state. In particular, it does
not require any knowledge of the explicit form of the extended Massieu
function. We sketch here the proof originally given by Katz
in~\cite{katzI}. It has been reformulated for systems like BHs with
several control parameters in~\cite{Okamoto:1994aq}. Other proofs have
been given by Sorkin~\cite{Sorkin:1981jc,Sorkin:1982ut}.

Assume that an expression for the extended Massieu function is
given, at least in the neighbourhood of ${\cal M}_{\rm eq}$. We can
parametrize such a function as:
\begin{equation}
  \label{OffShellSParametrization}
  \widehat{\cal S}=\widehat{\cal S}(X^\rho;\mu^i)\, ,
\end{equation}
where the $\{X^\rho\}$ denote a set of off-equilibrium variables.
Equilibria take place at the values $X_{\rm eq}^\rho=X_{\rm
eq}^\rho(\mu^i)$ which are solutions of:
\begin{equation}
  \label{EquationsForEquilibria}
  \partial_\rho\widehat{\cal S} \equiv 
  \frac{\partial\widehat{\cal S}}{\partial X^\rho} = 0\, . 
\end{equation}
The equilibrium Massieu function~(\ref{GeneralFundamentalRelation}) is
then given by ${\cal S}_{\rm eq}(\mu^i)=\widehat{\cal S}(X_{\rm
eq}^\rho(\mu^i);\mu^i)$. Define off-equilibrium conjugate variables
$\widehat\beta_i$,
\begin{equation}
  \label{OffEquilibriumTemperatures}
  \widehat{\beta}_i(X^\rho;\mu^j)\equiv \partial_i\widehat{\cal S} 
  \equiv \frac{\partial\widehat{\cal S}}{\partial\mu^i}\, .
\end{equation}
These give the equilibrium values of
Eq.~(\ref{GeneralEquationsOfState}) when evaluated at equilibrium,
that is, $\beta_i(\mu^j)=\widehat{\beta}_i(X_{\rm
eq}^\rho(\mu^j);\mu^j)$. The point $X_{\rm eq}^\rho$ represents a
stable equilibrium if and only if it is a local maximum of
$\widehat{\cal S}$ at fixed $\{\mu^i\}$.  Therefore, one has to
consider the eigenvalues of the matrix
\begin{equation}
  \left.
  \partial_{\rho\sigma}\widehat{\cal S}(X^\tau;\mu^i)
  \right|_{X^\tau=X_{\rm eq}^\tau(\mu^i)}. 
\end{equation}
A {\em change} of stability along $X_{\rm eq}^\rho(\mu^i)$ takes place
when one of these eigenvalues becomes zero and changes sign. Assume
now that the variables $\{X^\rho\}$ can be chosen in such a way that
the above matrix is diagonal. The eigenvalues we are interested in are
then:
\begin{equation}
  \lambda_\rho (\mu^i) \equiv \left. 
  \partial_{\rho}^2\widehat{\cal S}(X^\sigma;\mu^i)
  \right|_{X^\sigma=X_{\rm eq}^\sigma(\mu^i)}\, .
\end{equation}
These are called ``Poincar\'e coefficients of stability''. An
instability appears whenever there is some $\lambda_\rho>0$. What we
will show now is how it is possible to get information about the {\em
sign} of an eigenvalue near a point where it vanishes {\em without}
having to compute the spectrum of $(\partial_{\rho\sigma}\widehat{\cal
S})_{\rm eq}$.

Consider the identities provided by
equations~(\ref{EquationsForEquilibria}) when the l.h.s. is evaluated
at $X_{\rm eq}^\rho$. Choose some $\mu^a$ (for fixed $a$) and take the
partial derivative with respect to it at both sides
of~(\ref{EquationsForEquilibria}). Because of the choice of normal
coordinates, one gets the set of relations:
\begin{equation}
  \label{KatzIdentity1}
  (\partial_\rho\widehat{\beta}_a)_{\rm eq} +
  \lambda_\rho\, \partial_aX^\rho_{\rm eq} = 0\, ,
\end{equation}
where there is no sum over $\rho$ in the second term. Now consider
Eq.~(\ref{OffEquilibriumTemperatures}) for $\widehat{\beta}_a$ (with
the same $a$ as before) also evaluated at $X_{\rm eq}^\rho$. Taking
again the derivative w.r.t. $\mu^a$ gives:
\begin{equation}
  \label{KatzIdentity2}
  \frac{\partial\beta_a}{\partial\mu^a} =
  (\partial_a^2\widehat{\cal S})_{\rm eq} +
  \sum_\rho{(\partial_\rho\widehat{\beta}_a)_{\rm eq}\, 
  \partial_aX^\rho_{\rm eq}}\, 
\end{equation}
(no sum over $a$). The important thing is that the l.h.s. contains
only the derivative of the equilibrium equation of
state. Using~(\ref{KatzIdentity1}) we get:
\begin{equation}
  \label{EqSlopesAgainstEigenvalues}
  \frac{\partial\beta_a}{\partial\mu^a} =
  (\partial_a^2\widehat{\cal S})_{\rm eq} - \sum_{\rho}
  {\frac{(\partial_\rho\widehat{\beta}_a)_{\rm eq}^2}
        {\lambda_\rho}}\, .
\end{equation}
Assume for a moment that at a generic equilibrium
$(\partial_a^2\widehat{\cal S})_{\rm eq}$ is finite and that the
coefficients $(\partial_\rho\widehat{\beta}_a)_{\rm eq}$ are finite
and nonzero (the failure of these assumptions will be discussed
below). In such a case, the last equation shows that when a change of
stability occurs (i.e. when one or several eigenvalues $\lambda_\rho$
become zero and change sign) the plot of the conjugacy diagram
$\beta_{a}(\mu^a)$ {\em along the equilibrium series} has a vertical
tangent, since $\partial\beta_a/\partial\mu^a$ has to diverge
there~---~this is a ``turning point'' (see Fig.~\ref{fig:BetaMu}). The
first consequence of this is that:
\begin{itemize}
  \item[(i)] If one single point of an equilibrium sequence is shown
  to be fully stable, then all equilibria in the sequence are fully
  stable until the first turning point is reached.
\end{itemize}
Let us consider the general case in which one can have degeneracy of
zeromodes at a given equilibrium state, and let
$\tilde\rho=1,\cdots,N$ label the eigenvalues becoming zero. Near the
turning point one has:
\begin{equation}
  \label{SlopeNearATP} 
  \frac{\partial\beta_a}{\partial\mu^a} \approx -
  \sum_{\tilde\rho}{ 
  \frac{(\partial_{\tilde\rho}\widehat{\beta}_a)_{\rm eq}^2}
  {\lambda_{\tilde\rho}}} + \cdots \ {\rm (finite\ terms)}\, .
\end{equation}
This shows that:
\begin{itemize}
  \item[(ii)] At a turning point in the diagram $\beta_a(\mu^a)$, the
  branch with negative slope {\em near} the turning point is always
  unstable,
\end{itemize}
since a negative slope near the turning point means that at least one
term in the r.h.s. of~(\ref{SlopeNearATP}) has to be negative.

In a generic case, however, we cannot conclude that the branch with
positive slope near the turning point is stable: for example, there
can always be positive eigenvalues that do not change sign and, in
such a case, they will never show up in a conjugacy diagram. So the
positive slope branch is not guaranteed to be fully stable unless we
do know that at least one point in the branch is in fact fully
stable. On the other hand:
\begin{itemize}
  \item[(iii)] If the spectrum of zeromodes is nondegenerate, the
  degree of stability changes by one at a turning point. The branch
  with positive slope near the turning point is always more stable
  than the one with negative slope.
\end{itemize}
Here ``more stable'' means that it has exactly one negative mode less
than the negative slope branch, since a nondegenerate spectrum means
that no more than one eigenvalue is changing sign at a time. The
positive slope branch will be stable against the particular eigenvalue
that changes sign at the turning point and, in that sense, it is
``more stable''.

In any case, what one can always see with this method is the existence
of {\em instabilities}: the existence of a turning point from which a
negative slope branch emerges necessarily implies the existence of at
least one unstable mode. Point (ii) above asserts that this conclusion
remains true even without the assumption of a nondegenerate spectrum
of zeromodes and without any knowledge of stability of any particular
point in the sequence~\cite{Sorkin:1981jc,Sorkin:1982ut}.

\subsubsection*{Bifurcations}

However, it is still possible to have an eigenvalue
$\lambda_{\tilde{\rho}}$ that changes sign but whose behaviour does
not show up as a turning point in the conjugacy diagrams
$\beta_a(\mu^a)$. This can only happen if, at the point of change of
stability, the coefficient
$(\partial_{\tilde\rho}\widehat{\beta}_a)_{\rm eq}$ also
vanishes. Then we will not see in general any vertical tangent. It can
be shown that this can only happen at a {\em bifurcation}, i.e. when
there is another sequence of equilibria that crosses the considered
one, the intersection point being the bifurcation. One can indeed
prove that~\cite{Iooss}:
\begin{itemize}
  \item[(iv)] Changes of stability can {\em only} occur at turning
  points or bifurcations.
\end{itemize}

The general result is therefore that, in the absence of bifurcations,
the stability character of a sequence cannot change until the
equilibrium curve meets a turning point. In particular, stability does
not change when the slope of $\beta_a(\mu^a)$ changes sign when
passing through zero, thus ruling out the criteria based on the sign
of the specific heats (see e.g. Eq.~(\ref{Stab1})) for nonextensive
systems.  This is because, in general, the sign of the slope of the
diagram $\beta_a(\mu^a)$ does not say anything about the stability
character of any mode. It is only in the {\em vicinity} of a turning
point where both things can be related via Eq.~(\ref{SlopeNearATP}).

\subsubsection*{Vertical Asymptotes}

One might have wondered why, from the appearance of a zero eigenvalue
in~(\ref{EqSlopesAgainstEigenvalues}), we infer the existence of a
vertical {\em tangent}, rather than a vertical asymptote. The reason
is that at a vertical asymptote there are no equilibrium states at
all, and therefore they signal the (asymptotic) endpoint or
``boundary'' of the equilibrium sequence. In particular, one cannot
speak about any ``change of stability'' there, since if equilibria
exist beyond a vertical asymptote they belong to a different
sequence. 

On the other hand, it seems more reasonable to expect the divergence
at an asymptote to be a consequence of the non-differentiability of
$\widehat{\cal S}$ at the boundary of the sequence, rather than to an
eigenvalue approaching zero. The 4 dimensional Kerr BH discussed below
is an example where indeed one can prove that the divergence at a
vertical asymptote has to be due to some non-differentiability of
$\widehat{\cal S}$. We will consider in detail vertical asymptotes in
the cases studied below, and we will see that there are important
differences w.r.t. the behaviour at turning points.

In the following we will admit non-smoothness of $\widehat{\cal S}$
only at such points. If one does not have at least continuous second
order derivatives then thermodynamic coefficients do not exist, and
the Gaussian approximation to the entropy breaks down.

\subsubsection*{Examples}

\begin{figure}[!ht]
\begin{center}
\leavevmode
\epsfxsize= 8cm
\epsffile{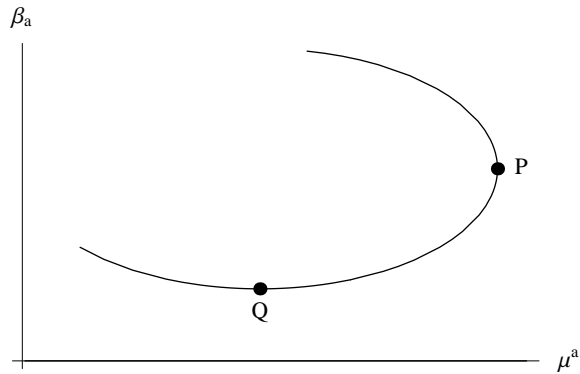}
\caption{\footnotesize{Generic plot of a conjugate variable $\beta_a$ 
against $\mu^a$ along an equilibrium sequence. The point $P$ is a
turning point, where a change of stability can take place. No changes
of stability occur at $Q$, even if the slope changes sign there.}}
\label{fig:BetaMu}
\end{center}
\end{figure}

A typical conjugacy diagram is shown in Fig.~\ref{fig:BetaMu}. The
point $P$ is a turning point.  The upper branch is unstable, while the
lower branch can be stable or more stable. The sign of the slope also
changes at the horizontal tangent at $Q$, but this has no relation
with a change of stability. Also, if somehow one knows that there are
no bifurcations, then one can conclude that the stability of such a
curve can change only at $P$. This means that the lower branch would
preserve its stability character beyond $Q$, even if the slope changes
sign there.

Let us note that here is a simpler reason to discard the point $Q$ as
a point of stability change along the equilibrium series. When
crossing $Q$, the control parameter $\mu^a$ does not remain constant,
so stability considerations based on maximization of $\widehat{\cal
S}$ simply do not apply there, since $\widehat{\cal S}$ is maximum
{\em at constant} $\mu^i$ (note, on the other hand, that $d\mu^a=0$ at
$P$). This was essentially the reason invoked in~\cite{Concave} to
disregard horizontal tangents in conjugacy diagrams.
\newline

\begin{figure}[!ht]
\begin{center}
\leavevmode
\caption{\footnotesize{Conjugacy diagrams for the four dimensional Kerr 
black hole~\cite{Kaburaki:TSKerrBH}.}}
\mbox{\subfigure[$\beta(M)$ at fixed $J$. $M_{\rm ext}$ stands for the 
                 minimal mass at fixed $J$ (the Kerr bound).
		 $M\rightarrow\infty$ is the 
                 Schwarzschild limit. $M_D$ is the ``Davies' point''
                 $J_D^2=(2\sqrt{3}-3)M_D^4$~\cite{Davies:1978mf}.]
                 {\epsfxsize=8cm \epsffile{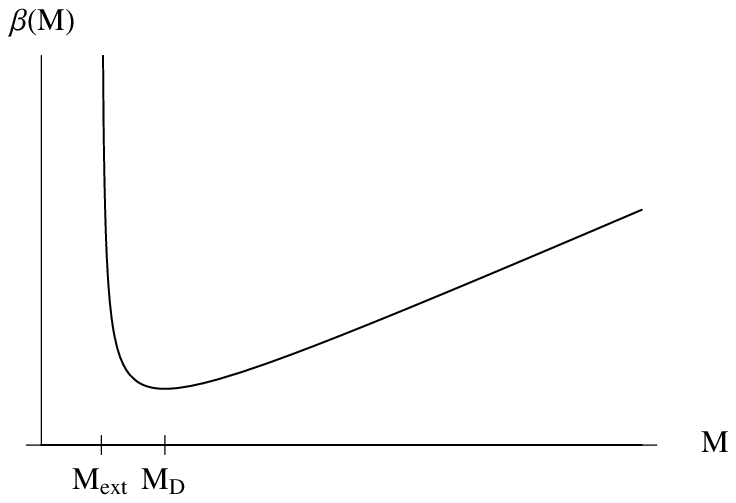}}
                 \subfigure[$\omega(J)$ at fixed mass.  $J_{\rm ext}$
                 stands for the Kerr bound. As in the plot 
		 of $\beta(M)$, one finds a vertical asymptote there.]  
                 {\epsfxsize=8cm \epsffile{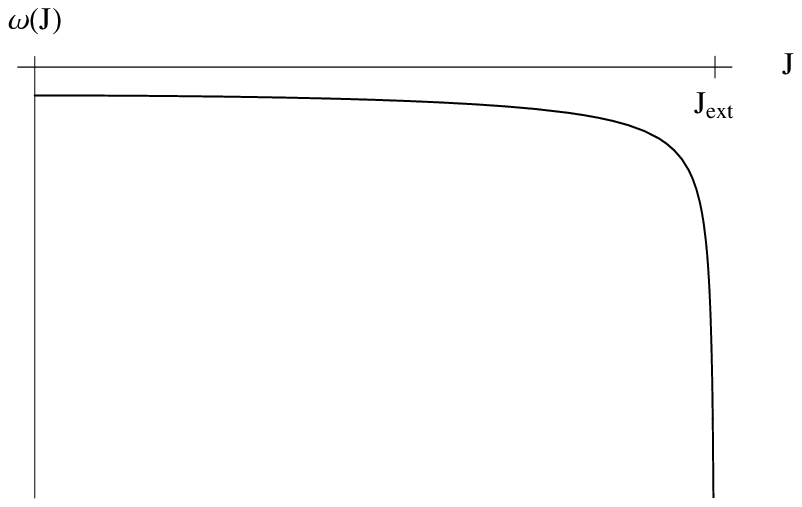}}}
\label{fig:Kerr4d}
\end{center}
\end{figure}

As a concrete example to make the previous arguments explicit,
consider the four dimensional Kerr black hole in the microcanonical
ensemble~\cite{Kaburaki:TSKerrBH}. The Massieu function is the entropy
$S$ and the control parameters are $\{\mu^i\}=\{M,J\}$. The conjugate
thermodynamic coordinates are $\{\beta_i\}=\{\beta,\omega\}$, $\beta$
being the inverse temperature and $\omega=-\Omega/T$.  For the study
of stability we need the plots of $\beta(M)$ at constant $J$ and
$\omega(J)$ at constant $M$ along the equilibrium series. Such plots
are given in Fig.~\ref{fig:Kerr4d}. \newline

The minimum at $M_D$ in the plot of $\beta(M)$ is exactly the point
discussed by Davies~\cite{Davies:1978mf}. There the specific heat at
constant $J$ changes sign through an infinite discontinuity, a fact
that has been interpreted in the literature as a sign of a phase
transition. However, this point is a horizontal tangent, and therefore
no changes of stability take place at $M_D$. Actually, these plots
show that there are no turning points at all. Therefore, according to
this analysis, no changes of stability occur in all the way down from
the Schwarzschild limit to extremality. There could be bifurcations,
but in this case these are ruled out due to the uniqueness of time
independent BH solutions of the Einstein equations in four
dimensions. Since we know that one point of the curve (namely, the
Schwarzschild limit) is stable~\cite{Kay:1987ax}, we can conclude that
stability is preserved along the whole curve, at least against the
dynamical modes accounted for in this analysis (see the discussion
below). The non-appearance of instabilities in these diagrams is a
result which clearly agrees nicely with the fact that Kerr black holes
are actually stable~\cite{Teukolsky:1973ha}.

One should consider the origin of the divergence at the vertical
asymptote at the Kerr bound in view of of
Eq.~(\ref{EqSlopesAgainstEigenvalues}). If the divergence at
extremality was only due to some $\lambda_{\tilde\rho}$ approaching
zero, then that eigenvalue had to be positive (i.e. unstable), since
the slope is negative. The absence of turning points in these curves
would imply that this unstable mode never changes sign along the whole
series, thus meaning that every point is unstable. This is impossible
due to the fact that one point of the curve, the Schwarzschild limit,
is stable. So the divergence at extremality (or, at least, the leading
contribution to it) has to be due to the non-differentiability of
$\widehat{S}$ at the Kerr bound. This makes sense, since the extremal
black hole has zero temperature and $\beta$ and $\omega$ themselves
diverge there. So even without knowing about the behaviour of
$\widehat{S}$, one should expect a divergence due to the ``zero
temperature contribution''.

In this case one can say even more about the behaviour of
$\widehat{S}$ near the vertical asymptote. In principle, both
coefficients $(\partial_a^2\widehat{\cal S})_{\rm eq}$ and
$(\partial_\rho\widehat{\beta}_a)_{\rm eq}$
in~(\ref{EqSlopesAgainstEigenvalues}) can diverge. However, given the
fact that we know that all the eigenvalues $\lambda_\rho$ have to stay
negative near extremality, the fact that $\partial\beta/\partial\mu^a$
is negative there implies that, near extremality:
\begin{equation}
  \label{SlopeNearAVA}
  \frac{\partial\beta_a}{\partial\mu^a} \approx 
  (\partial_a^2\widehat{S})_{\rm eq} + \cdots \ 
  {\rm (subleading\ terms)} . 
\end{equation}
Note that the above approximation means that the behaviour of the
phase diagrams $\beta_a(\mu^a)$ is controlled by completely different
physics at a vertical asymptote and at a turning point (compare
to~(\ref{SlopeNearATP})). This will be important in
Section~\ref{sec:FCBAndTG} when discussing fluctuations and trying to
isolate the relevant degrees of freedom both at turning points
and at vertical asymptotes.

\subsection{Relation to Dynamical Stability}
\label{sec:RelationToDS}

One should ask about a precise relation between the study of stability
based on the Poincar\'e method and the dynamical stability under
linear perturbations of the black hole solution. There are two issues
one has to consider at this point: the choice of the ensemble and the
number and nature of fluctuations.

\subsubsection*{Choice of the Ensemble}

We will be interested in the stability of solutions like~(\ref{MPBH})
and~(\ref{BRxy}). These describe isolated BHs, for which the mass and
angular momentum are conserved quantities that remain fixed under
Lorentzian time evolution. The appropriate ensemble is thus the
microcanonical, and the relevant potential is the entropy.
Inequivalence of the stability limits of different ensembles in
self-gravitating systems has been discussed
in~\cite{Kaburaki:TSKerrBH,Katz:1993up,KabuKerr,Concave,Parentani:1994wr}
and, from a more general point of view, in~\cite{Parentani:1994aa}.
This becomes clear from the precedent discussion on the existence of
turning points: a different ensemble means a different choice of the
Massieu function and, typically, a reciprocal choice of the conjugate
variables, which may in general interchange horizontal tangents
(``Davies' points'') by vertical ones (turning points). This is what
would happen if, for example, we decide to put the above discussed
$d=4$ Kerr black hole in the canonical ensemble. In that case
stability would be linked to the maximization of the ``free energy''
$\phi=S-\beta M$. The conjugacy diagram of interest would be
$M=M(\beta)$, and such a plot has a vertical tangent where
$\beta=\beta(M)$ happens to have a horizontal one. Therefore we see,
as mentioned at the beginning of this Section, that the point $M_D$
only means a change of stability of the black hole when put in contact
with a thermal bath, but not of an isolated BH.

\subsubsection*{Nature of Fluctuations}

As discussed above, the appearance of a single unstable mode at a
turning point is enough to ensure instability of a black hole
solution. However, we are still faced with the question of which {\em
dynamical} (metric) fluctuations are accounted for in a conjugacy
diagram. Put differently, we would like to know which deformations of
the metric carry physically the off-equilibrium fluctuations denoted
above by $\{X^\rho\}$. In general one will have some identification of
the kind
\begin{displaymath}
  X^\rho\sim\delta g_{\mu\nu}^{(\rho)},
\end{displaymath}
where the $\{\delta g_{\mu\nu}^{(\rho)}\}$ stand for a suitable set of
metric perturbations. This has to be so because there are no other
possibilities for coordinates in the configuration space\footnote{For
an explicit derivation in specific cases of these kind of
identifications from a gravitational path-integral point of view see
e.g.~\cite{Parentani:1994wr,Katz:2000vv} and references therein.}. The
problem is to identify which particular perturbations are encoded in
our analysis.

This question is important because, in principle, we do not have
enough arguments to conclude that full Poincar\'e stability (if
achieved) is enough to guarantee full dynamical stability. First it
seems difficult that in a general case the conjugate thermodynamic
variables could encode all possible physical metric perturbations. In
addition, note that applicability of the turning point method requires
some assumptions on the spectrum of
$(\partial_{\rho\sigma}\widehat{\cal S})_{\rm eq}$: namely, a discrete
spectrum and the possibility of choosing normal coordinates for the
off-equilibrium variables. Without any knowledge about the
off-equilibrium function $\widehat{\cal S}$, and in the lack of a
rigorous proof, one has to admit the possibility that these
requirements may be, in a generic case, too strong to be satisfied by
the whole spectrum of $(\partial_{\rho\sigma}\widehat{\cal S})_{\rm
\rm eq}$ under arbitrary linearized perturbations. There could be some
perturbations that do not satisfy the required conditions for them to
be ``scanned'' with this method, and some of those could be unstable
modes.

Perturbations satisfying the above criteria are probably those keeping
some degree of symmetry\footnote{On physical grounds, it is also
unlikely that asymmetric perturbations will extremize the entropy (see
e.g.~\cite{SWZ81}). In such a case, the most natural possibility is
that the system meets another branch of less-symmetric equilibria
(i.e. a bifurcation).}. In~\cite{Kaburaki:TSKerrBH,Katz:1993up} it has
been argued that, for rotating BHs, Poincar\'e stability ensures
stability against axisymmetric perturbations\footnote{See
also~\cite{Friedman:1988er} for another example concerning the
symmetry characterizing the relevant perturbations.}. It would be nice
to have an explicit proof of this. \\

Therefore we see that the turning point method seems reliable when
detecting the appearance of instabilities. Moreover, one of its main
advantages is its extreme simplicity and, most importantly, the fact
that it can be applied to non-extensive systems such as BHs. The
drawbacks of this method are, however, the following: 1)~Bifurcations
do not show up in the conjugacy diagrams. 2)~Branches with a positive
slope at a turning point are not ensured to be fully stable unless one
can prove stability of at least one point in the branch. 3)~If it
cannot be proven that the spectrum of zeromodes is nondegenerate, then
one has no control on the relative number of negative modes between
two branches meeting at a turning point. 4)~In principle, there might
be dynamical modes that this treatment of fluctuations does not
encode.

Before ending this Section, let us mention that this stability
analysis only requires from the considered system an evolution
dictated by the extremization of some potential. In our case, such a
potential is the BH area, which we know that obeys (classically) an
extremum law due to Hawking's area theorem. In particular, application
of this method does not require interpretation of the black hole area
or surface gravity as a thermodynamic entropy or temperature.

\newpage
\subsection{Stability of Black Holes and Black Rings}

Let us now apply these considerations on stability to the five
dimensional black hole/black ring system. Both for the sake of
comparison and for later use, let us turn first to the study of the
Hessian of the entropy.

\subsubsection{Standard Thermodynamical Stability}

\begin{figure}[!ht]
\begin{center}
\leavevmode
\epsfxsize= 8cm
\epsffile{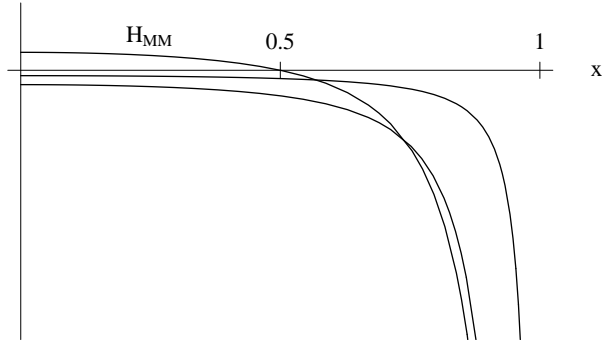}
\caption{\footnotesize{$H_{MM}(x)$, $H_{JJ}(x)$ and $\det H(x)$ 
for the rotating BH. The curve changing sign at $x=1/2$ is that of the
element $H_{MM}$. $H_{JJ}$ and $\det H$ are always negative.}}
\label{fig:HessianBH}
\end{center}
\end{figure}

We denote the elements of the Hessian by
$H_{ij}=\partial^2S/\partial\mu^i\partial\mu^j$, where the variables
$\{\mu^i\}$ are $\{M,J\}$. For the rotating BH the fundamental
relation that we need is given by~(\ref{BHEntropy}). In terms of the
variable $x$ defined in~(\ref{x}) and the mass parameter $\mu$ one
finds that (from now on we set $G=1$):
\begin{equation}
  \label{HessianBH}
  \left\{
  \begin{array}{rclcl}
  H_{MM} &=&
  \displaystyle
  \frac{2^3}{3\sqrt{\mu}}\,
  \frac{1-4x^2}{(1-x^2)^{3/2}} \ \ \ \ \ \ \ \ 
  &\sim&\displaystyle
  -\frac{2\sqrt{2}}{\sqrt{\mu}}\,
  \frac{1}{\epsilon^{3/2}} + 
  \frac{23\sqrt{2}}{6\sqrt{\mu}}\,
  \frac{1}{\epsilon^{1/2}} + 
  {\cal O}(\epsilon^{1/2})\, , 
  \\[.4cm]
  H_{MJ} &=&
  \displaystyle
    \frac{2^3}{\mu}\,
    \frac{x}{(1-x^2)^{3/2}}
  &\sim&\displaystyle
  +\frac{2\sqrt{2}}{\mu}\, 
  \frac{1}{\epsilon^{3/2}} -
  \frac{\sqrt{2}}{2\mu}\, 
  \frac{1}{\epsilon^{1/2}} + 
  {\cal O}(\epsilon^{1/2})\, , 
  \\[.4cm]
  H_{JJ} &=&
  \displaystyle
  - \frac{2^3}{\mu^{3/2}}\,
    \frac{1}{(1-x^2)^{3/2}}
  &\sim&\displaystyle
  -\frac{2\sqrt{2}}{\mu^{3/2}}\, 
  \frac{1}{\epsilon^{3/2}} -
  \frac{3\sqrt{2}}{2\mu^{3/2}}\, 
  \frac{1}{\epsilon^{1/2}} + 
  {\cal O}(\epsilon^{1/2})\, , 
  \\[.4cm]
  \det H &=&
  \displaystyle
  - \frac{2^6}{3\mu^2}\,
    \frac{1}{(1-x^2)^2}
  &\sim&\displaystyle
  -\frac{2^4}{3\mu^2}\, 
  \frac{1}{\epsilon^2} - 
  \frac{2^4}{3\mu^2}\, 
  \frac{1}{\epsilon} + 
  {\cal O}(1)\, ,
  \end{array}
  \right.
\end{equation}
where in the r.h.s. we have quoted the divergent part in the expansion
around $\epsilon\equiv 1-x$, which we shall need later. The plots of
the diagonal elements and the determinant at fixed mass are given in
Fig.~\ref{fig:HessianBH}.

The element $H_{MM}$ is proportional to minus the inverse specific
heat $c_J$ at constant $J$, so one can see that $c_J$ changes sign
from negative to positive values through an infinite discontinuity at
$x=0.5$, exactly as in the four dimensional case. Therefore the same
comments about what happens at $x=0.5$ apply here (see
Section~\ref{sec:BHsAndNES}). On the other hand one would conclude,
according to the standard criteria, that the system is unstable for
all values of the angular momentum, since the Hessian always has one
positive eigenvalue. \newline

As for the BRs, the fundamental relation that we need is given
by~(\ref{BRFundamentalRelation}). In terms of the parameter $\nu$, the
elements of the Hessian are given by:
\begin{equation}
  \label{HessianBR}
  \left\{
  \begin{array}{rcl}
  H_{MM} &=& \displaystyle
  -\frac{2^4\sqrt{2}}{3\sqrt{\mu}}\,
  \frac{\sqrt{\nu}\, (\nu^2+2)}{(2\nu-1)\, (1-\nu)^{3/2}}\, , \\[.4cm]
  H_{MJ} &=& \displaystyle
  +\frac{2^5}{\mu}\,
  \frac{\nu}{\sqrt{1+\nu}\, (2\nu-1)\, (1-\nu)^{3/2}}\, , \\[.4cm]
  H_{JJ} &=& \displaystyle
  -\frac{2^6\sqrt{2}}{\mu^{3/2}}\,
  \frac{\nu^{3/2}}{(1+\nu)^2\, (2\nu-1)\, (1-\nu)^{3/2}}\, , \\[.4cm]
  \det H &=& \displaystyle
  -\frac{2^{10}}{3\mu^2}\,
  \frac{\nu^2}{(1+\nu)^2\, (2\nu-1)\, (1-\nu)^2}\, ,
  \end{array}
  \right.
\end{equation}
where one only has to replace $\nu$ by the appropriate value
$\nu_S(x)$ or $\nu_L(x)$ in~(\ref{nuLS}). Recall that $\nu=1/2$ gives
the ring with lowest angular momentum, $\nu=0$ represents the infinite
angular momentum limit of the large black ring, and $\nu=1$ is the
$x=1$ limit of the small black ring. For later use, let us write down
the divergent part of the expansions around $\delta\equiv 2\nu-1$ and
$\epsilon\equiv 1-\nu$:
\begin{equation}
  \label{HessianBRExpansion}
  \left\{
  \begin{array}{rclcl}
  H_{MM}& \sim & \displaystyle
		 -\frac{2^33\sqrt{2}}{\sqrt{\mu}}\, 
                 \frac{1}{\delta} + {\cal O}(1)\, , \ \ & 
  \sim & \displaystyle
		 -\frac{2^4\sqrt{2}}{\sqrt{\mu}}\, 
                 \frac{1}{\epsilon^{3/2}} - 
                 \frac{2^35\sqrt{2}}{3\sqrt{\mu}}\, 
                 \frac{1}{\epsilon^{1/2}} + 
                 {\cal O}(\epsilon^{1/2})\, , \\[.4cm]
  H_{MJ}& \sim & \displaystyle
		 +\frac{2^6\sqrt{3}}{3\mu}\, 
                 \frac{1}{\delta} + {\cal O}(1)\, , \ \ & 
  \sim & \displaystyle
		 +\frac{2^4\sqrt{2}}{\mu}\, 
                 \frac{1}{\epsilon^{3/2}} + 
                 \frac{2^25\sqrt{2}}{\mu}\, 
                 \frac{1}{\epsilon^{1/2}} + 
                 {\cal O}(\epsilon^{1/2})\, , \\[.4cm]
  H_{JJ}& \sim & \displaystyle
		 -\frac{2^8\sqrt{2}}{3^2\mu^{3/2}}\, 
                 \frac{1}{\delta} + {\cal O}(1)\, , \ \ & 
  \sim & \displaystyle
		 -\frac{2^4\sqrt{2}}{\mu^{3/2}}\, 
                 \frac{1}{\epsilon^{3/2}} - 
                 \frac{2^33\sqrt{2}}{\mu^{3/2}}\, 
                 \frac{1}{\epsilon^{1/2}} + 
                 {\cal O}(\epsilon^{1/2})\, , \\[.4cm]
  \det H& \sim & \displaystyle
		 -\frac{2^{12}}{3^3\mu^2}\, 
                 \frac{1}{\delta} + {\cal O}(1)\, , \ \ & 
  \sim & \displaystyle
		 -\frac{2^8}{3\mu^2}\, 
                 \frac{1}{\epsilon^2} - 
                 \frac{2^8}{3\mu^2}\, 
                 \frac{1}{\epsilon} + 
                 {\cal O}(1)\, .
  \end{array}
  \right.
\end{equation}

\begin{figure}[!ht]
\begin{center}
\leavevmode
\caption{\footnotesize{Plots of $H_{MM}$, $H_{JJ}$ and $\det H$
as a function of $x$ for both black rings.}}
\mbox{\subfigure[$H_{MM}(x)$, $H_{JJ}(x)$ and $\det H(x)$
                 for the small black ring. All quantities diverge
                 both at the lower bound $x_{\rm min}$ and at the
                 upper one $x=1$.]
                 {\epsfxsize=7.5cm \epsffile{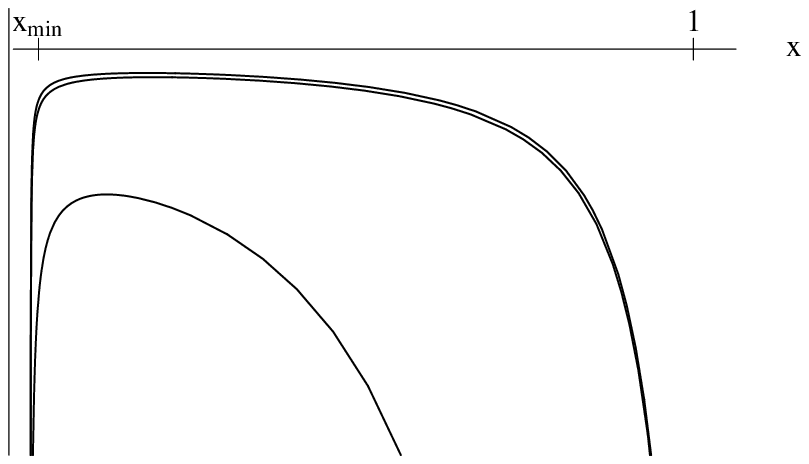}}
      \subfigure[$H_{MM}(x)$, $H_{JJ}(x)$ and $\det H(x)$
                 for the large black ring. All quantities diverge
                 at the lower bound $x_{\rm min}$ and tend
                 asymptotically to zero at $x\rightarrow \infty$.]
                 {\epsfxsize=7.5cm \epsffile{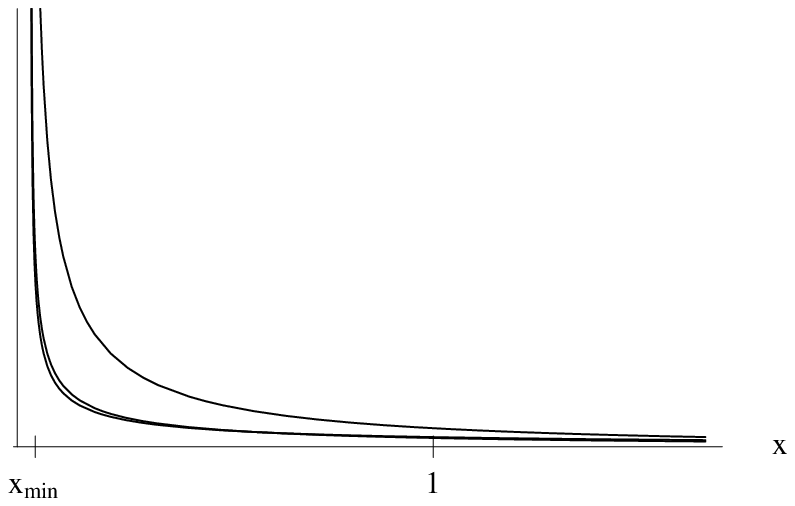}}}
\label{fig:BRs-Sigmas}
\end{center}
\end{figure}

Fig.~\ref{fig:BRs-Sigmas} is a plot of the different quantities at
fixed mass for both the small and the large black ring.
Again, we can see that for both configurations there is always a
positive eigenvalue of the Hessian: the SBR has one positive
eigenvalue, while the LBR has two positive eigenvalues for all $x$.
The divergence of the small black ring at $x=1$ was expected, since
this solution approaches the BH solution in the extremal limit (see
e.g.  Fig.~\ref{fig:Entropies}). Notice the infinite discontinuity at
$x_{\rm min}$ between the small and large black ring cases, even if
both solutions coincide and are nonsingular at $x_{\rm min}$.

\subsubsection{Turning Points and Stability Analysis}

\begin{figure}[!ht]
\begin{center}
\leavevmode
\caption{\footnotesize{The conjugacy diagrams $\beta(M)$ and $\omega(J)$ 
for all phases of the BH/BR system along their corresponding
equilibrium series. The dashed lines represent the BH-curves. We see
the existence of a turning point in both diagrams at the point where
both BR branches meet. The SBR and BH curves have a vertical asymptote
at extremality. $\beta_{\rm BH}(M)$ has a minimum (not shown in the
plot) at the value of $M$ that corresponds to $x=1/2$.}}
\mbox{\subfigure[$\beta(M)$ at fixed $J$. $M_{\rm ext}$ stands 
                 for the minimal allowed mass at fixed angular
                 momentum in the BH and SBR cases (the ``Kerr'' or
                 extremal bound $x=1$). $M_{\rm min}$ is the mass at 
                 $x_{\rm min}$.]
                 {\epsfxsize=8cm \epsffile{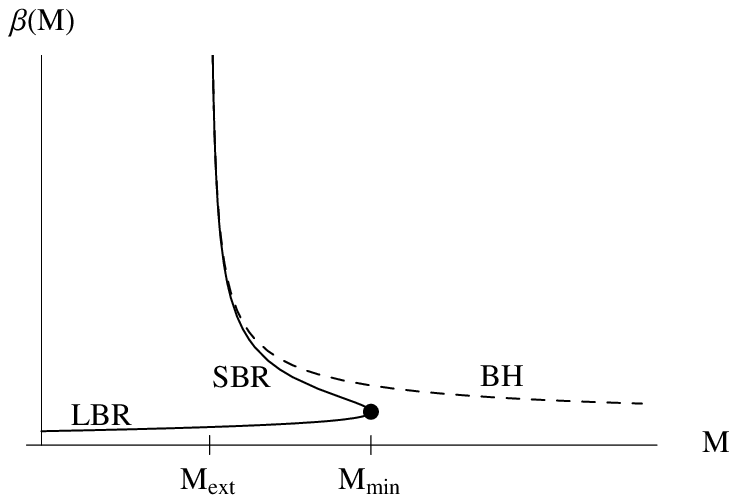}}
      \subfigure[$\omega(J)$ at fixed mass.]  
                 {\epsfxsize=8cm \epsffile{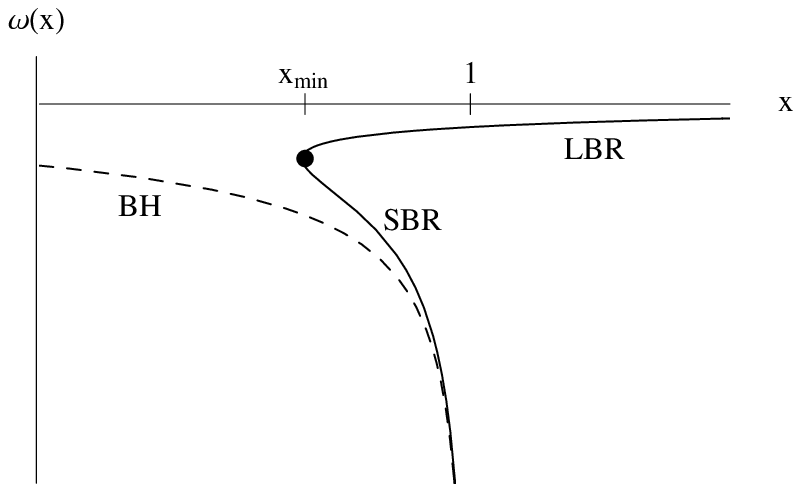}}}
\label{fig:StabilityBHBR}
\end{center}
\end{figure}

To perform the Poincar\'e analysis and investigate the existence of
turning points we need now the plots of $\beta(M)$ at fixed angular
momentum and $\omega(J)$ ($\omega\equiv-\Omega/T$) at fixed mass. For
the black hole and black rings, the conjugate variables are
respectively given by:
\begin{equation}
  \label{betaomegaBHBR}
  \begin{array}{cc}
  \left\{
  \begin{array}{rcl}
    \beta_{\rm BH}  & = & \displaystyle
    \frac{2\pi\, \sqrt{\mu}}{\sqrt{1-x^2}}\, , \\[.4cm]
    \omega_{\rm BH} & = & \displaystyle
    - \frac{2\pi\, x}{\sqrt{1-x^2}}\, . 
  \end{array}
  \right. & \ \ \ \ \ \ \ 
  \left\{
  \begin{array}{rcl}
    \beta_{\rm BR}  & = & \displaystyle
    2\pi\sqrt{2}\sqrt{\mu}\,
    \sqrt{\frac{\nu}{1-\nu}}\, , \\[.4cm]
    \omega_{\rm BR} & = & \displaystyle
    -\frac{4\pi\, \nu}{\sqrt{1-\nu^2}}\, .
  \end{array}
  \right.
   \end{array}
\end{equation}
All these curves, for the BH and BR phases, are represented in
Fig.~\ref{fig:StabilityBHBR}. \newline

Focusing first on the BH curves (dashed lines) we see that there are
no turning points, and therefore no changes of stability are shown
along the black hole series. Given the fact that the five dimensional
Schwarzschild solution is stable~\cite{Ishibashi:2003ap}, we can
conclude that {\em the five dimensional Myers-Perry BH with just one
angular momentum is stable, unless:}
\begin{itemize}
  \item[a)] there are bifurcations, which would imply non-uniqueness
  of the BH solution {\em other than black rings}, or 
  \item[b)] one admits the possibility of the existence of dynamical
  modes that the turning point method does not ``see''.
\end{itemize}
With respect to point~a) above, let us point out that the point
$x\approx 0.943$ in Fig.~\ref{fig:Entropies} is not a bifurcation,
since the BH and the LBR are different spacetimes. One can check that
the function $\beta_{\rm BH}=\beta_{\rm BH}(M)$ has a minimum
(horizontal tangent) at $M=27\pi J^2/8$ (i.e. $x=1/2$) which is not
shown in the figure\footnote{The whole BH curves are analogous to
those of Fig.~\ref{fig:Kerr4d}.}. This is the ``Davies' point'' where
the specific heat changes sign but, as already discussed, no changes
of stability occur there. On the other hand, the extremal limit is a
vertical asymptote, again like in the four dimensional case. The five
dimensional case is completely analogous in all respects, and
therefore we can equally conclude that the divergence at $x=1$ comes
from the non-differentiability of $\widehat{S}$, and that expressions
like~(\ref{SlopeNearAVA}) also apply here.\\

Before proceeding to the analysis of the BR curves, let us derive some
general properties which, assuming no bifurcations, will be of use
when analyzing the degeneracy of the spectrum of zeromodes. The
general expressions of the kind of~(\ref{EqSlopesAgainstEigenvalues})
are, in this case:
\begin{equation}
  \label{KatzForAllHij}
  \left\{
  \begin{array}{rcl}
  \partial_M\beta & = & 
  \displaystyle
  (\partial_{MM}\widehat{S})_{\rm eq} 
  - \sum_{\sigma}
  {\frac{(\partial_\sigma\widehat{\beta})_{\rm eq}^2}
        {\lambda_\sigma}}
  - \sum_{\tilde\rho}
  {\frac{(\partial_{\tilde\rho}\widehat{\beta})_{\rm eq}^2}
        {\lambda_{\tilde\rho}}}\, , \\[.4cm]
  \partial_J\omega & = &
  \displaystyle
  (\partial_{JJ}\widehat{S})_{\rm eq} 
  - \sum_{\sigma}
  {\frac{(\partial_\sigma\widehat{\omega})_{\rm eq}^2}
        {\lambda_\sigma}}
  - \sum_{\tilde\rho}
  {\frac{(\partial_{\tilde\rho}\widehat{\omega})_{\rm eq}^2}
        {\lambda_{\tilde\rho}}}\, , \\[.4cm] 
  \partial_{MJ}S & = &
  \displaystyle
  (\partial_{MJ}\widehat{S})_{\rm eq} 
  - \sum_{\sigma}
  {\frac{(\partial_\sigma\widehat{\beta})_{\rm eq}
         (\partial_\sigma\widehat{\omega})_{\rm eq}}
        {\lambda_\sigma}}
  - \sum_{\tilde\rho}
  {\frac{(\partial_{\tilde\rho}\widehat{\beta})_{\rm eq}
         (\partial_{\tilde\rho}\widehat{\omega})_{\rm eq}}
        {\lambda_{\tilde\rho}}}\, .
  \end{array}  
  \right.
\end{equation}
These are the elements of the equilibrium Hessian. The off-diagonal
term ($=\partial_M\omega=\partial_J\beta$) can be computed in a
similar fashion as we did in Section~(\ref{sec:BHsAndNES}) to compute
the diagonal ones. In the equations above, sum over $\sigma$ denotes
sum over never-zero eigenvalues, while sum over $\tilde\rho$ denotes
the sum over all eigenvalues becoming zero at some point. Note first
that, near such a point, the diagonal elements of the Hessian have to
diverge all with the same power:
\begin{equation}
  \label{HiiBehaviour}
  H_{MM} \sim H_{JJ} \sim \frac{1}{\lambda_{\tilde{1}}} 
  + \cdots \ {\rm (subleading\ terms)}\, , 
\end{equation}
$\lambda_{\tilde{1}}$ denoting the eigenvalue(s) contributing with the
highest power to the sum over $\tilde{\rho}$. Notice that this need
not be the case if the divergence was not due to the appearance of a
zeromode. On the other hand, the leading contribution to the
determinant of the Hessian is given by:
\begin{equation}
  \label{LeadingDivergence}
  \det H \sim 
  \left(\sum_{\tilde\rho}
  {\frac{(\partial_{\tilde\rho}\widehat{\beta})_{\rm eq}^2}
        {\lambda_{\tilde\rho}}}\right)
  \left(\sum_{\tilde\rho}
  {\frac{(\partial_{\tilde\rho}\widehat{\omega})_{\rm eq}^2}
        {\lambda_{\tilde\rho}}}\right)-
  \left(\sum_{\tilde\rho}
  {\frac{(\partial_{\tilde\rho}\widehat{\beta})_{\rm eq}
         (\partial_{\tilde\rho}\widehat{\omega})_{\rm eq}}
        {\lambda_{\tilde\rho}}}\right)^2\, + \cdots \ 
  {\rm (s.t.)}\, ,
\end{equation}
thus meaning that, if there are $N>1$ zeromodes
$\lambda_{\tilde\rho}$, then the leading contribution to the
determinant near a turning point will in general go like:
\begin{equation}
  \det H \sim \frac{1}{\lambda_{\tilde{1}}\lambda_{\tilde{2}}} 
  + \cdots\, \ {\rm (s.t.)}, 
\end{equation}
where $\lambda_{\tilde{2}}$ represents the eigenvalue with the highest
subleading contribution to the divergence of $H_{ii}$
in~(\ref{HiiBehaviour})\footnote{In the case where
$\lambda_{\tilde{1}}$ is degenerate, the divergence will go
like~$\sim 1/\lambda_{\tilde{1}}^2$.}. If, on the contrary, there is no
degeneracy of zeromodes, then one will always have the nontrivial
property:
\begin{equation}
  \label{NonDegOrder}
  H_{MM} \sim H_{JJ} \sim H_{MJ} \sim \det H 
  \sim \frac{1}{\lambda_{\tilde{1}}} + 
  \cdots\ {\rm (s.t.)}\, ,
\end{equation}
since the leading divergence in~(\ref{LeadingDivergence}) will exactly
vanish, and the contribution of order~$1/\lambda_{\tilde{1}}$ to the
off-diagonal term cannot be cancelled. Moreover, in view
of~(\ref{KatzForAllHij}), the numerical coefficients $h_{ij}^{(1)}$ of
the leading contributions to the divergence of the elements $H_{ij}$
have to satisfy the exact relation:
\begin{equation}
 \label{NonDegExact}
 \left(h_{MJ}^{(1)}\right)^2 = h_{MM}^{(1)}\ h_{JJ}^{(1)}\, .
\end{equation}
This remains true order by order in degree of divergence also in the
case of a degenerate spectrum in which all zeromodes contribute with a
different degree of divergence. \newline

Let us turn now to the analysis of stability of black rings. From the
conjugacy diagrams we see that there are two special points:
$x_{\min}$ and $x=1$. At $x_{\rm min}$ we find a turning point. Taking
into account the general results of the preceding Section, we
therefore conclude that {\em the small black ring is always
unstable}. From the entropy diagram in Fig.~\ref{fig:Entropies} we
knew that the small black ring was {\em globally} unstable, since it
is the phase with lower entropy. Now we see that it is unstable also
locally.

What about the stability of the large black ring? First we see, from
the behaviour of the Hessian near the turning
point~(\ref{HessianBRExpansion}), that there is only one divergent
piece and, in addition, that the necessary
conditions~(\ref{NonDegOrder}) and~(\ref{NonDegExact}) for a
nondegenerate spectrum are satisfied. Therefore this provides strong
evidence that {\em the large black ring is more stable} (in degree
one) {\em than the small ring}. However, as discussed above, we cannot
infer from this fact its full dynamical stability. Actually, one
expects on physical grounds that at least the limit of large $x$ may
suffer from a Gregory-Laflamme instability: the large angular momentum
limit describes a very long and thin ring, and so it may be considered
as a ``black string limit'' of the
configuration~\cite{Emparan:2004wy,EssayRoberto}. If there are GL
instabilities in the LBR, there are three possibilities to explain the
absence of turning points in the conjugacy diagrams.
\begin{itemize}
  \item[a)] The first one is that these diagrams are ``blind'' against
  GL modes. This would be the case if, as hinted at
  in~\cite{Kaburaki:TSKerrBH,Katz:1993up}, the dynamical modes
  accounted for in this analysis are only the ones preserving
  axisymmetry. Notice that instabilities of the Gregory-Laflamme type
  tend to distort the shape of the string along its axis. In our case,
  this means deformations along the $S^1$ within the plane of
  rotation, and therefore they are not axisymmetric perturbations.
  \item[b)] The second possibility is that the GL instability is
  present for all values of $x$. In this case the LBR would be always
  unstable but there would be no turning points. 
  \item[c)] The last possibility is that the emergence of GL modes
  actually appears as a bifurcation. This would imply the existence of
  a new branch of solutions beyond some angular momentum per unit
  mass, most probably describing non-uniform black
  rings~\cite{Horowitz:2001cz,Gubser:2001ac}. We consider this latter
  possibility an interesting one which deserves further investigation.
\end{itemize}
A further heuristic argument, due to Emparan and Reall, in favour of
another instability of the large and small black rings is the
following. As discussed in~\cite{EssayRoberto}, one expects the limit
$x=x_{\min}$ to be unstable: one simply has to throw matter with no
angular momentum into the ring with minimal $x$. This would decrease
the value of the angular momentum per unit mass of the final state
and, given the absence of black ring states at $x<x_{\rm min}$, such a
process would destabilize the system. This could happen if
possibilities analogous to~a) or~b) above apply to such an
instability. \newline

Let us comment now on the origin of the divergence at the vertical
asymptote that the SBR curves exhibit at $x=1$. As discussed for the
black hole case, it seems most likely that such a divergence appears
due to the fact that $\beta$ and $\omega$ themselves diverge there
because of the zero temperature at $x=1$. In the SBR case, however,
one has to admit the possibility that the divergence can be due to an
unstable mode that approaches zero at extremality since, contrary to
the BH, the SBR has negative modes for all $x$. We want to argue here
that this does not seem the case. The BH and SBR solutions approach
each other as they get close to $x=1$, so one should expect the
physical origin of this divergence to be the same in both cases. In
fact, by looking at the leading contributions ($\sim\epsilon^{-3/2}$)
to the divergence in the expansion of the SBR
Hessian~(\ref{HessianBRExpansion}) at $x=1$, we see that these exactly
coincide with those of the BH~(\ref{HessianBH}) up to an overall
constant factor. So we find evidence that the divergence at $x=1$, or
at least the leading contribution $\sim\epsilon^{-3/2}$, is due to the
fact that the temperature vanishes there. Therefore, as in the four
dimensional case, we will write:
\begin{equation}
  \label{LeadingHessianAtExtremality}
  H_{ij} \approx (\partial_{ij}\widehat{S})_{\rm eq} 
  + \cdots \ {\rm (subleading\ terms)}\, ,
\end{equation}
both for the BH and the SBR near extremality (we are assuming that
this expression also holds for the off diagonal term
of~(\ref{KatzForAllHij})). This will be of importance in the next
Section, when we will consider the behaviour of fluctuations. \newline

Summarizing, our analysis proves {\em local} instability of the SBR
and presents strong evidence that the LBR is a ``more stable'' phase,
even though it does not guarantee its full dynamical stability. In
fact, one expects the LBR to suffer from instabilities which do not
show up in the conjugacy diagrams, the possible reasons for this being
the ones discussed above. On the other hand, these conjugacy diagrams
do not show any special property neither at $x=1/2$ (``Davies'
point'') nor at $x=2\sqrt{2}/3$ (where the LBR becomes entropically
favoured, see Fig.~\ref{fig:Entropies}).


\section{Fluctuations, Critical Behaviour and\\ Thermodynamic Geometry}
\label{sec:FCBAndTG}

When an unstable point is reached one expects fluctuations to diverge,
as it also happens for instance at the critical point of a second
order phase transition. However, both phenomena are not the same. Let
us next elaborate on this issue in some detail.

Before proceeding we want to make clear what we mean by
``fluctuations'' in the context of the off-equilibrium formalism
introduced above. What we want to consider here are general motions in
configuration space, i.e. motions parametrized by $\{\Delta
X^\rho,\Delta\mu^i\}$. Note that for the study of {\em stability} we
needed to work at fixed $\{\mu^i\}$, as required by entropy
maximization. Here, however, we will not be interested in stability
considerations, but rather in the changes induced in the system as we
move in {\em all} possible directions of configuration space. This is
why now we will consider variations like $\{\Delta\mu^i\}$ as well.

The physical significance of the variations in the $\{X^\rho\}$- and
in the $\{\mu^i\}$-directions is, however, different. The
off-equilibrium variables $\{X^\rho\}$ should be considered as
``dynamical'', in the sense that they obey a first order equilibrium
equation ($\partial_\rho\widehat{S}=0$). Therefore, the variations
$\{\Delta X^\rho\}$ can be thought as spontaneous fluctuations of the
system. On the other hand, the set of variables $\{\mu^i\}$ are our
control parameters (in the microcanonical ensemble, the mass and the
angular momentum) and, as such, they do {\em not} fluctuate
spontaneously. In any case, we can always consider the changes induced
in the system as we change these control parameters
(``adiabatically'', say), in an attempt to map out the behaviour of
the system (or better, the thermodynamic potential) in configuration
space at the vicinity of equilibrium\footnote{Note that variations
$\Delta\mu^i$ will not in general run along the equilibrium series
unless they are correlated with the $\Delta X^{\rho}$ as dictated by
the equilibrium equation $X^{\rho}_{\rm eq}=X^{\rho}(\mu^i)$.}. At
this point, let us recall that all our considerations are purely
classical, and therefore we are ignoring any quantum effects like
Hawking radiation (in such a case, spontaneous variations in mass and
angular momentum would be of course physical).

\subsection{Fluctuations at Turning Points and Vertical Asymptotes}
\label{sec:DFAndCPh}

Let us fisrt consider fluctuations around equilibria near a turning
point, following in part the discussions
in~\cite{Parentani:1994wr,Okamoto:1994aq,Katz:2000vv}. If the
eigenvalue $\lambda_{\tilde{1}}$ becoming zero there is nondegenerate,
then the fluctuations will be basically driven by fluctuations in the
corresponding mode $X^{\tilde{1}}$, since this is the direction along
which the Massieu potential becomes flat, the restoring forces are
small and then $\Delta X^{\tilde{1}}$ can be large. This will induce
off-equilibrium fluctuations in the conjugate variables
$\widehat{\beta}_i$, which will be well approximated by:
\begin{equation}
  \label{TemperatureFluctuations}
  \delta\widehat{\beta}_i\approx
  (\partial_{\tilde{1}}\widehat{\beta}_i)_{\rm eq}
  \Delta X^{\tilde{1}}\,  
\end{equation}
(unless $(\partial_{\tilde{1}}\widehat{\beta}_i)_{\rm eq}$ is
vanishingly small, which can only happen at a bifurcation --- see
Section~\ref{sec:BHsAndNES}).  This approximation remains valid also
if there are other zeromodes, but still the one approaching faster to
zero is nondegenerate.

The probability for the system to be at $X=X_{\rm eq}+\Delta
X^{\tilde{1}}$ is proportional to the exponential of the Massieu
function evaluated at $X$~\cite{Callen,ll}, which will be well
approximated by its second order variation along that direction:
\begin{equation}
  \label{EntropyFluctuations}
  P(X_{\rm eq}+\Delta X^{\tilde{1}}) 
  \sim \exp\left[\widehat{\cal S}_{\rm eq} + 
  \textstyle\frac{1}{2}
  \lambda_{\tilde{1}} (\Delta X^{\tilde{1}})^2\right]\, .
\end{equation}
The last two expressions together with Eq.~(\ref{SlopeNearATP}) show
that, {\em near a turning point} which is not a bifurcation and where
$\lambda_{\tilde{1}}$ is nondegenerate, the probability density for a
fluctuation $\delta\widehat{\beta}_i$ is given by
\begin{equation}
 P(\delta\widehat{\beta}_i) \sim 
 \exp\left[ -\frac{1}{2}\, 
     \frac{(\delta\widehat{\beta}_i)^2}
          {\partial\beta_i/\partial\mu^i}\right]\, .    
\end{equation}
Calculating the normalization factor and evaluating the mean square
deviation (from the side with $\lambda_{\tilde{1}}<0$) is
straightforward, and gives the
result~\cite{Parentani:1994wr,Okamoto:1994aq,Katz:2000vv}:
\begin{equation}
  \label{MSDNearTP}
  \langle(\delta\widehat{\beta}_i)^2\rangle \approx 
  \frac{\partial\beta_i}{\partial\mu^i}\, . 
\end{equation}
The same result was obtained in~\cite{Concave,KabuKerr} via somewhat
different considerations. There the $\delta\widehat{\beta}_i$ were
taken from the start as good coordinates to parametrize
off-equilibrium states, thus casting all the discussion on
off-equilibrium fluctuations directly in terms of them (without
passing through the $\{X^\rho\}$ coordinates used here). In such a
case, the above formula, which here is obtained only as an
approximation near a turning point, turns out to be valid
everywhere\footnote{This can be seen from the general formalism by
substituting in~(\ref{EntropyFluctuations}) (which in this case holds
for all modes, which are precisely the $\delta\widehat{\beta}_i$) the
exact value of $(\partial_{\tilde{1}}^2\widehat{\cal S})_{\rm eq}$ as
obtained from Eq.~(\ref{EqSlopesAgainstEigenvalues}) and taking each
$\widehat{\beta}_i$ as independent from the $\{\mu^i\}$ and the rest
of the $X^\rho$ ($\sim \widehat{\beta}_j$, $j\neq i$).}. However, in
general one will need a different and more complete parametrization to
account for all off-equilibrium states. In such a case the above
equality is ensured to hold only near a turning point.

So we see that, when the system approaches a turning point from the
stable side, the quadratic fluctuations of the conjugate variables
diverge. In the BH/BR system this happens at $x_{\rm min}$, where we
also expect the condition of $\lambda_{\tilde{1}}$ being nondegenerate
to be satisfied (see last Section). In such a case we can write, near
$x_{\rm min}$ (see~(\ref{KatzForAllHij})):
\begin{equation}
  H_{MM} \approx
           -\frac{(\partial_{\tilde{1}}\widehat{\beta})^2}
           {\lambda_{\tilde{1}}}\, , \ \ \ 
  H_{JJ} \approx 
           -\frac{(\partial_{\tilde{1}}\widehat{\omega})^2}
           {\lambda_{\tilde{1}}}\, , \ \ \ 
  H_{MJ} \approx 
           -\frac{(\partial_{\tilde{1}}\widehat{\beta})
                  (\partial_{\tilde{1}}\widehat{\omega})}
           {\lambda_{\tilde{1}}}\, .
\end{equation}
Let us recall that the dominant term in the expansion of the BR
Hessian~({\ref{HessianBRExpansion}}) exactly
obeys~(\ref{NonDegExact}), i.e. the numerical coefficients are in
perfect agreement with the approximations above. Calculating the ratio
$\delta\widehat{\beta}/\delta\widehat{\omega}$
from~(\ref{TemperatureFluctuations}) allows to write,
from~(\ref{MSDNearTP}) and $H_{MJ}$ above, the off-diagonal
correlations $\langle\delta\widehat{\beta}\,
\delta\widehat{\omega}\rangle$.  We thus obtain the general
expression:
\begin{equation}
  \label{dbidbjTP}
  \langle\delta\widehat{\beta}_i\, \delta\widehat{\beta}_j\rangle
  \approx + H_{ij}\,  
\end{equation}
for correlations near a turning point. The expression above is then
telling us how fluctuations of the LBR diverge as the system
approaches $x_{\rm min}$.\newline

Let us turn now to the fluctuations in the BH and the SBR cases near
the vertical asymptote at $x=1$. Here the phenomenon is very
different, since to start with is not related to a change of
stability. We saw in the last Section that the leading divergence of
the elements of the Hessian can be explained by the divergence of the
elements $(\partial_{ij}\widehat{\cal S})_{\rm
eq}$. Eq.~(\ref{LeadingHessianAtExtremality}) implies that these
coefficients are the ones that dominate also the series expansion of
the fluctuations $\delta\widehat{\beta}$ and $\delta\widehat{\omega}$
(since it implies that $(\partial_{ij}\widehat{\cal S})_{\rm eq}\gg$
$(\partial_{\rho}\widehat{\beta}_i)_{\rm eq}$
$(\partial_{\rho}\widehat{\beta}_j)_{\rm eq}$) and, in addition, that
they are well approximated by the equilibrium Hessian. Therefore we
can write:
\begin{equation}
  \label{DeltaBetaExtremality}
  \delta\widehat{\beta}_i \approx H_{ij}\Delta\mu^j\, .
\end{equation}
So we see that, contrary to the previous case (see
Eq.~(\ref{TemperatureFluctuations})), the variations
$\delta\widehat{\beta}_i$ are now dominated by the variations in the
variables $\{\mu^i\}$, which are small but still induce the main
contribution to $\delta\widehat{\beta}_i$ because of the large
coefficient. Variations in the $\{\mu^i\}$-directions will be governed by
a probability density which,
using~(\ref{LeadingHessianAtExtremality}), can be written as
approximately proportional to
\begin{equation}
  \label{DeltaSExtremality}
  P(\mu_{\rm eq}^i+\Delta\mu^i) \sim \exp\left[\widehat{S}_{\rm eq} +
  \beta_i\Delta\mu^i + 
  \frac{1}{2}H_{ij}\Delta\mu^i\Delta\mu^j\right]\, .
\end{equation}
Expressions~(\ref{DeltaBetaExtremality}) and~(\ref{DeltaSExtremality})
formally coincide with those used in the usual treatments of
fluctuations in thermodynamic fluctuation theory\footnote{Up to the
linear term in the probability distribution, whose effect just amounts
to a shift of the mean values $\langle\delta\widehat{\beta}_i\rangle$
(see Appendix). As shown below, the contribution of this term to the
mean square deviations is subleading. The linear term is nonvanishing
here because the $\{\mu^i\}$ do not obey any ``equilibrium equation''
(i.e. the vanishing of a first order variation) --- only the
$\{X^\rho\}$ variables do.}~\cite{ll}. The computation of correlations
gives the result:
\begin{equation}
  \label{dbidbjVA}
  \langle\delta\widehat{\beta}_i\, \delta\widehat{\beta}_j\rangle
  \approx - H_{ij} + \beta_i\beta_j \approx -H_{ij}\, 
\end{equation}
(note the minus in the final expression sign w.r.t.~(\ref{dbidbjTP})).
The computation of these fluctuations requires some care, however, due
to the presence of a positive eigenvalue in the Hessian matrix. Its
derivation is outlined in the Appendix. The first equality is the
exact result as computed from the (approximate) probability
distribution~(\ref{DeltaSExtremality}). The additional contribution to
the Hessian is due to the presence of the linear term in the
probability distribution. However, such a contribution can be
neglected near extremality, since it is subleading as
$\epsilon\rightarrow 0$ (see Eqns.~(\ref{HessianBH}),
(\ref{HessianBRExpansion}) and (\ref{betaomegaBHBR})). 

Summarizing, we see that both near extremality and near the turning
point fluctuations diverge, and that at those points correlations can
be computed from the equilibrium Hessian. \newline

Another situation where fluctuations typically diverge is at a
critical point. However, as pointed out in~\cite{Concave,kabuscaling},
this is in general a different phenomenon from a change of
stability. A critical point is the endpoint of a ``coexistence
curve'', characterized by the existence of two competing local maxima
of the corresponding Massieu function, describing for example the
liquid and gas phases of a fluid. At the critical point both maxima
coalesce, which entails a locally flat potential and an absence of
restoring forces. This is what makes fluctuations become large near
the critical point. On the other hand, the divergence of fluctuations
related to a change of stability is in general due to a different
mechanism: the flattening of the potential because the system enters
into a series of local minima (unstable equilibria) from a series of
local maxima (stable equilibria).

Next we would like to investigate these phenomena in the BH/BR
system. To this aim, further information on the critical behaviour of
a system can be obtained from the geometry of the thermodynamic state
space, which is what we shall introduce next. \newline

Before ending this Section the following comment is in order. Note
that, at least in the SBR case (and probably in the LBR also), we are
considering fluctuations of an {\em unstable} system (i.e. with some
$\lambda_{\rho}>0$) . Therefore, to make sense of all approximations
above, we have been forced to suppress by hand all variations along
the unstable $\{X^{\rho}\}$-directions, which will be of course the
dynamically relevant ones. All our considerations are thus purely
formal, in the sense that we are restricting ourselves to study these
configurations along the directions in configuration space in which no
unstable modes are present. It is in this sense in which this analysis
on fluctuations (and the subsequent discussion on a phase transition)
has to be understood.

\subsection{Critical Phenomena and the Geometry of the\\ 
Thermodynamic State Space}
\label{sec:CriticalAndRuppeiner}

It has been observed by several authors in different contexts that the
entropy function induces a natural metric on the thermodynamic state
space, and that the geometric invariants constructed out of it may
provide interesting information about the phases of the model under
consideration. This formalism was pioneered by Ruppeiner, and we refer
the reader to~\cite{Rupp} for a very detailed review and various
examples. Ruppeiner formalism relies on the usual thermodynamic
properties of extensive systems. Here we want to argue that, modulo
some careful reinterpretations, it can be applied also to BHs {\em
near turning points and vertical asymptotes}.  Previous works where
Ruppeiner and related thermodynamic geometries have been applied to
black hole physics include~\cite{Ferrara:1997tw}, but the issue of the
non-extensivity of BHs has not been discussed there.

Let us first review how this formalism can be motivated in ordinary
(extensive) thermodynamics~\cite{Rupp}. Let us consider the
microcanonical ensemble, in which one starts with a homogeneous,
isolated system in the thermodynamic limit. The purpose is to
investigate thermodynamic fluctuations in a given open subsystem (to
be thought as a ``small'' subsystem) of fixed volume $V$. The
microcanonical variables $\{\mu^i\}$ are fixed within the total
system, but the ones specifying the state of the subsystem
$\{\mu^i_V\}$ can fluctuate around their equilibrium values
$\left\{\mu_{V\, {\rm eq}}^i\right\}$, which will be set by those of
the environment. All off-equilibrium states of the subsystem are
assumed to be uniquely specified by the values of the $\{\mu^i_V\}$.

According to the basic assumption of thermodynamic fluctuation theory,
the probability density $P_V$ of finding the subsystem in a state
characterized by values $\{\mu_V^i\}$ of the remaining variables of
state (all except the volume, which is fixed) is given by:
\begin{equation}
  \label{ProbDensitySubsystem}
  P_V(\mu_V^i)
  \sim \exp\left[S_{\rm tot}(\mu_V^i;\mu_{\rm eq}^i)\right]\, ,
\end{equation}
where the $S_{\rm tot}$ is the entropy of the total system and
$\mu_{\rm eq}^i$ are the average (equilibrium) values of the
environment. The function $S_{\rm tot}(\mu_V^i;\mu_{\rm eq}^i)$ has,
as a function of $\{\mu_V^i\}$, a local maximum when the densities of
the subsystem equal those of the environment, i.e. at
$\mu_V^i=\mu^i_{V\, {\rm eq}}\approx V\mu_{\rm eq}^i/V_{\rm tot}$. The
above formula describes fluctuations of the subsystem around this
equilibrium value. If one now assumes extensivity, i.e. the fact that
the entropy of the total system $S_{\rm tot}$ is the sum of that of
the subsystem $S_V$ plus that of the environment, and expands around
the equilibrium point up to second order, one gets~\cite{ll,Rupp}:
\begin{equation}
  \label{StandardGaussianExpansion}
  P_V(\mu_V^i) = \sqrt{\frac{\det g(\mu_{\rm eq})}{(2\pi)^{n-1}}}
  \exp\left[-\textstyle\frac{1}{2}
  g_{ij}(\mu_{\rm eq})\Delta\mu_V^i\Delta_V^j\right]\, ,
\end{equation}
where $(n-1)$ is the number of independent fluctuating variables once
the volume has been fixed. In the expression above $g_{ij}$ is 
defined as:
\begin{equation}
  \label{RuppMetricSV} g_{ij} (\mu_V) \equiv
  -\frac{\partial^2S_V}{\partial\mu_V^i\partial\mu_V^j}\, ,
\end{equation}
and $\Delta\mu_V^i=\mu_V^i-\mu_{V\, {\rm eq}}^i$. Sum over repeated
indices ($i,j=1,2,\cdots,n-1$) is understood in the argument of the
exponential.  In standard treatments the formula above is written in
terms of the entropy and other quantities per volume (i.e. $S_V=Vs$,
etc). Here we prefer to leave it in terms of the absolute quantities
of the subsystem itself to make closer contact with the black hole
case.

So far this is standard thermodynamic fluctuation theory~\cite{ll}.
Ruppeiner proposed to interpret the Hessian of the entropy $g_{ij}$ as
a metric in the thermodynamic state space.  This metric is normally
known in the statistical mechanics literature as the ``Ruppeiner
metric'', and it is supposed to govern the fluctuations of the
system. The meaning of distance as defined by this metric is the
following: the less the probability of a fluctuation between two
states, the further apart they are w.r.t. the geodesic distance
defined by this metric. In other words, the metric is an indicator of
the ``closeness'' of probability distributions. A straightforward
computation based on the Gaussian distribution shows that the average
values of fluctuations vanish:
\begin{equation}
  \label{AverageDeviationsMu}
  \langle\Delta\mu_V^i\rangle=0\, ,
\end{equation}
and that the main square deviations are given by the contravariant
metric tensor:
\begin{equation}
  \label{SquareDeviationsMu}
  \langle\Delta\mu_V^i\Delta\mu_V^j\rangle = g^{ij}(\mu_{\rm eq})\, ,
\end{equation}
which becomes an alternative definition of the thermodynamic metric.

Ruppeiner theory is {\em covariant}, in the sense that the line
element built from the metric defined in~(\ref{RuppMetricSV}) is taken
to be an invariant quantity. This prescription serves to tell us what
to do if the chosen coordinates to describe fluctuations of the
subsystem are not the usual extensive quantities. If for example we
choose to work with the conjugate variables in the entropy
representation
\begin{equation}
  \beta_{V i} = \frac{\partial S_V}{\partial\mu_V^i}\, , 
\end{equation}
the line element is written as:
\begin{equation}
  \label{LineElementPhi}
  ds^2 = g_{ij}(\mu_V)\, d\mu_V^id\mu_V^j = 
         -d\beta_{V i}d\mu_V^i =
         \frac{\partial^2\phi(\beta_V)}
              {\partial\beta_{V i}\partial\beta_{V j}}\, 
         d\beta_{V i}d\beta_{V j}\, ,
\end{equation}
where the new Massieu function $\phi$ is the Legendre transform of the
entropy w.r.t. the variables $\{\beta_{V i}\}$:
\begin{equation}
  \label{MetricElementsBeta}
  \phi(\beta_V) = S_V - \mu_V^i\beta_{V i}\, .
\end{equation}
Equation~(\ref{LineElementPhi}) gives the expression for the metric in
the new coordinates\footnote{Our use of upper and lower indices could
be confusing here. Notice however that a change of variables to the
$\{\beta_i\}$ makes them become the new genuine contravariant
coordinates. Expression~(\ref{MetricElementsBeta}) gives therefore the
{\em covariant} components of the metric in the new coordinates.},
\begin{equation}
  \tilde{g}_{ij}(\beta_V) \equiv \frac{\partial^2\phi(\beta_V)}
  {\partial\beta_{V i}\partial\beta_{V j}}\, .
\end{equation}
Expressions like~(\ref{AverageDeviationsMu})
and~(\ref{SquareDeviationsMu}) regarding fluctuations hold in any
chosen coordination of the state space.

\subsubsection*{Thermodynamic Curvature}

A very special quantity in Ruppeiner theory is the curvature computed
from the thermodynamic metric. For a two dimensional system ($n-1=2$),
as will be later our case of interest, the curvature scalar contains
all nontrivial information about the metric itself (in higher
dimensional cases one may need other components of the curvature
tensor).

We will not review here the derivations leading to the meaning of the
thermodynamic curvature scalar. We will simply recall its main
properties, referring the interested reader to the
Ref.~\cite{Rupp}. First, computation of the scalar curvature for known
statistical models provides indication that it is a measure of
interactions. For example, it is zero for the ideal gas, but the
thermodynamic metric turns out to be curved in the Van der Waals
case. But, most importantly for our purposes, the curvature diverges
at a critical point, and it does so in the same way as the correlation
volume $\xi^d$, i.e.
\begin{equation}
  \label{RuppAndCL}
  R_{\rm crit} \sim \xi^d\, ,
\end{equation}
where $\xi$ is the correlation length and $d$ the effective spatial
dimension of the system. The reason for this to be so can be put a bit
more precisely: calculations show~\cite{Rupp} that the scalar
curvature measures the smallest volume where one can describe the
considered subsystem as surrounded by a uniform environment. At
criticality, one expects such a volume to scale as $\xi^d$. 

\subsubsection*{Applicability to Black Holes}

So far we have sketched the main features of Ruppeiner formalism as
they can be motivated from ordinary, extensive thermodynamics. We
deduced the notion of a metric in the thermodynamic state space by
considering a subsystem of a much bigger extensive system in the
microcanonical ensemble. Such a total system effectively acts as a
reservoir for the subsystem, of fixed volume $V$. A careful derivation
of the theory just presented requires extensivity of the entropy
function. We have in fact invoked extensivity to arrive to the first
place where the thermodynamic metric appears,
Eq.~(\ref{StandardGaussianExpansion}). How can all this be conceivably
applied to non-extensive systems such as black holes?

We will try to argue now that this is in fact the case. All
considerations exposed so far come from the Gaussian
expansion~(\ref{StandardGaussianExpansion}). In statistical mechanics,
extensivity is required to infer the fundamental relation $S_V$ from
the knowledge of $S_{\rm tot}$, which is what is usually known. It is
also needed to ``split'' the degrees of freedom of the subsystem from
those of the reservoir. But note that, in practice, this splitting
only serves to fix the equilibrium values of the subsystem.  On the
other hand, since we are fixing one scale, the function $S_V$ will not
be an homogeneous function of its variables. In practice, what this
means is that we will be working with an entropy function $S_V$ which
is {\em not} extensive {\em within the subsystem}. The latter and its
fluctuations are the object of physical interest, and the formalism
described above does not meet in principle any difficulty as long as:
1)~the (non-homogeneous) function $S_V$ is known, independently of its
relation to $S_{\rm tot}$, and 2)~we have an expression for the
relevant off-equilibrium variations of the subsystem around its
(known) equilibrium state. This is precisely the situation we found
for black holes near turning points and vertical asymptotes, which
therefore have to be identified with ``Ruppeiner's subsystem''.

In Ruppeiner theory, general off-equilibrium states of the subsystem
are described exactly in Eq.~(\ref{StandardGaussianExpansion}) by the
values of the $\{\mu_V^i\}$. That is, all possible off-equilibrium
states of the subsystem are supposed to be well described by the
thermodynamic limit. In the the BH case, however, we do not have in
general any control on off-equilibrium fluctuations, since they are
given by the completely unknown
function~$\widehat{S}(X^{\rho};\mu^i)$. The analog of the Ruppeiner
metric (i.e. the coefficients of the quadratic expansion of the
entropy) will be given, in a general case, by the many-dimensional
metric with elements:
\begin{equation}
  g_{ij} = -(\partial_{ij}\widehat{S})_{\rm eq}\, , \ \ \ 
  g_{i\rho} = -(\partial_{i\rho}\widehat{S})_{\rm eq}\, , \ \ \ 
  g_{\rho\rho} = -\lambda_\rho\, . 
\end{equation}
However, {\em near a turning point or a vertical asymptote}, the
expressions~(\ref{EntropyFluctuations}) and~(\ref{DeltaSExtremality})
provide a good description of off-equilibrium fluctuations. Moreover,
the quadratic approximation to the entropy is computable, and the
``effective'' elements of the Ruppeiner metric are just a few and can
be computed from the equilibrium Hessian. In fact, one only has to use
the prescription~(\ref{SquareDeviationsMu}) for the inverse metric as
read from equations~(\ref{dbidbjTP}) and~(\ref{dbidbjVA}). 

\subsection{Thermodynamic Curvature for the BH and BR Spacetimes}

\begin{figure}[!ht]
\begin{center}
\leavevmode
\epsfxsize= 8cm
\epsffile{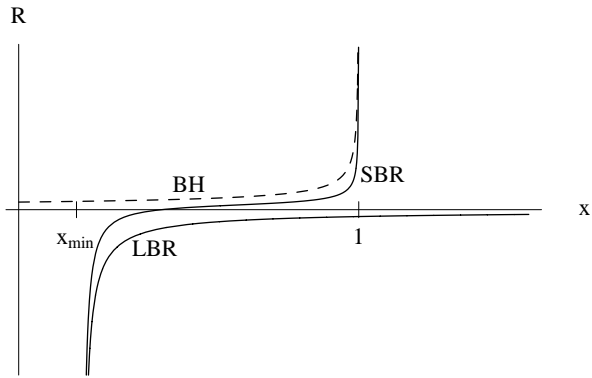}
\caption{\footnotesize{Thermodynamic curvature scalar for the BH 
(dashed line), SBR and LBR phases. The physically relevant regions are
those near $x_{\rm min}$ and $x=1$. The curvature is divergent at
those points.}}
\label{fig:RBHBR}
\end{center}
\end{figure}

As explained above, the elements of the Ruppeiner metric can be
computed from the expressions for quadratic fluctuations. Near $x_{\rm
min}$ and $x=1$ we will write, according to~(\ref{dbidbjTP})
and~(\ref{dbidbjVA})\footnote{For simplicity, we are neglecting the
subleading contribution $\beta_i\beta_j$ in~(\ref{dbidbjVA}). It can
be checked that this does not change the power behaviour of the
divergence of the thermodynamic curvature at extremality, which is
what we will be interested in.}:
\begin{equation}
  g^{ij} =
  \langle\delta\widehat{\beta}_i\, \delta\widehat{\beta_j}\rangle
  = \pm H_{ij}\, ,
\end{equation} 
where the plus sign stands for the LBR at $x_{\rm min}$ and the minus
sign for the SBR and BH at $x=1$. Let us compute now the curvature
scalar from these metric elements. For convenience, we give below the
expressions as computed from $g^{ij}=+H_{ij}$, and we will provide
their values for all $x$. The physically meaningful limits are however
those at $x_{\rm min}$ and $x=1$ with the appropriate sign. The
results for the rotating black hole and black rings are, respectively:
\begin{equation}
  \left\{
  \begin{array}{rcl}
    R_{\rm BH} & = & \displaystyle
    \frac{2}{\pi^2\mu^{3/2}} \frac{1}{\sqrt{1-x^2}}\, , \\[.4cm]  
    R_{\rm BR} & = & \displaystyle
    \frac{2\sqrt{2}}{\pi^2 \mu^{3/2}}\,
    \frac{\sqrt{\nu}\, (\nu^2+2\nu-2)}{(2\nu-1)^2 \sqrt{1-\nu}}\, .
  \end{array}
  \right.
\end{equation}
These quantities are plotted in Fig.~\ref{fig:RBHBR}. We see that
divergences of the curvature (i.e. indicators of critical behaviour
according to Ruppeiner theory) are found at $x_{\rm min}$ and $x=1$,
in agreement with our previous analysis on stability and fluctuations.


\section{Scaling Laws in the BH/BR System}
\label{sec:ScalingLawsBHBR}

Ruppeiner analysis confirms that the points $x_{\rm min}$ and $x=1$
are candidates of critical points. However, divergences in the
thermodynamic curvature have been found to happen also when the system
enters into an unphysical region. In~\cite{Brody:1999gc}, for
instance, computation of the thermodynamic scalar for the Van der
Waals model showed that it diverges also at the so-called spinoidal
curve. In our case, both at $x_{\rm min}$ and at $x=1$ the BH/BR
system enters in an unphysical region (namely, naked
singularities). One can also argue that the origin of these
divergences could just be the divergent nature of fluctuations at
those points (which, as discussed in Section~\ref{sec:DFAndCPh}, do
not necessarily mean critical behaviour). In order to clarify this
issue, we examine next the scaling relations between the different
critical exponents both at $x_{\rm min}$ and $x=1$.

\subsection{Critical Exponents and Order Parameter}

We shall first define the appropriate susceptibilities and critical
exponents that are suited to the microcanonical
ensemble. Following~\cite{kabuscaling} we define them as follows:
\begin{equation}
  \left\{
  \begin{array}{rclcl}
    \chi_J & \equiv & \displaystyle
    \frac{\partial^2S}{\partial M^2} & \sim &
      \epsilon_M^{-\alpha}\, , \ \ 
      \epsilon_J^{-\varphi}\, . \\[.4cm]
    \chi_M & \equiv & \displaystyle
    \frac{\partial^2S}{\partial J^2} & \sim &
      \epsilon_M^{-\gamma}\, , \ \  
      \epsilon_J^{-\sigma}\, .
  \end{array}
  \right.
\end{equation}
These are the natural generalizations of the expressions familiar from
other ensembles (for example, in the grand canonical ensemble the
thermodynamic potential is the Gibbs free energy, and the specific
heat is given by its second derivative with respect to the
temperature, etc). Also, in the expressions above the $\epsilon$
parameters are the suitable generalizations of what e.g. would be the
``reduced temperature'' $(T-T_{\rm crit})/T_{\rm crit}$ in the usual
treatments (i.e. (grand)canonical ensembles) of more familiar examples
like liquid/gas transitions, etc. We want to study the scaling
relations at the two points of the system where fluctuations diverge,
i.e. at $x_{\rm min}$ and at extremality, $x=1$. At each of these, we
define the parameters $\epsilon_M$ and $\epsilon_J$ to be:
\begin{equation}
  \begin{array}{ll}
  {\rm At \ } x_{\rm min} {\rm : \ \ }
  \left\{
  \begin{array}{rcl}
    \epsilon_M & = & \displaystyle
    \frac{M_{\rm min}-M}{M_{\rm min}}\, , \\[.4cm]
    \epsilon_J & = & \displaystyle
    \frac{J-J_{\rm min}}{J_{\rm min}}\, ,
  \end{array}
  \right. & \ \ \ \
  {\rm At \ } x=1 {\rm : \ \ }
  \left\{
  \begin{array}{rcl}
    \epsilon_M & = & \displaystyle
    \frac{M-M_{\rm ext}}{M_{\rm ext}}\, , \\[.4cm]
    \epsilon_J & = & \displaystyle
    \frac{J_{\rm ext}-J}{J_{\rm ext}}\, ,
  \end{array}
  \right.
  \end{array}
\end{equation}
where
\begin{equation}
  \begin{array}{ll}
  \left\{
  \begin{array}{rcl}
    M_{\rm min} & = & \displaystyle
    \left(\frac{\pi J^2}{G}\right)^{1/3}\, , \\[.4cm]
    J_{\rm min} & = & \displaystyle
    \sqrt{\frac{GM^3}{\pi}}\, ,
  \end{array}
  \right. & \ \ \ \ \ \ \ \
  \left\{
  \begin{array}{rcl}
    M_{\rm ext} & = & \displaystyle
    \left(\frac{27\pi J^2}{32G}\right)^{1/3}\, , \\[.4cm]
    J_{\rm ext} & = & \displaystyle
    \sqrt{\frac{32GM^3}{27\pi}}\, .
  \end{array}
  \right.
  \end{array}
\end{equation}

Other critical exponents tell us about the behavior of the {\em
order parameter} of the transition considered. In general, there
is no obvious choice for an order parameter. In~\cite{kabuscaling}
it was proposed to choose, as an order parameter for BH phase
transitions, the difference of the ``intensive'' (conjugate)
variables between the two phases. In our case this means:
\begin{equation}
  \left\{
  \begin{array}{rl}
  {\rm At \ } x=x_{\rm min} {\rm :} & \ \ \
  \eta \equiv \omega_{\rm SBR} - \omega_{\rm LBR}\, , \\[.4cm]
  {\rm At \ } x=1 {\rm :} & \ \ \
  \eta \equiv \omega_{\rm SBR} - \omega_{\rm BH}\, .
  \end{array}
  \right.
\end{equation}
These parameters go to zero at the (presumed) transition point, as
they should. For these order parameters, the natural generalizations
of the critical exponents $\beta$ and $\delta$ familiar from
statistical mechanics is given by:
\begin{equation}
  \eta \sim \epsilon_{M}^{\beta} \sim \epsilon_{J}^{1/\delta}\, .
\end{equation}
With these definitions, from equations~(\ref{HessianBH})
and~(\ref{HessianBR}) the critical exponents are readily computed to
be:
\begin{equation}
  \begin{array}{rl}
  {\rm At\ } x=x_{\rm min} {\rm :} &
  \left\{
  \begin{array}{ll}
  {\rm Small \ Black \ Ring:} &
  \alpha = \varphi = \gamma = \sigma = 1/2\, , \\[.4cm]
  {\rm Large \ Black \ Ring:} &
  \alpha = \varphi = \gamma = \sigma = 1/2\, , \\[.4cm]
  \beta = 1/\delta = 1/2\, . 
  \end{array}
  \right. \\[1.2cm]
  {\rm At\ } x=1 {\rm :} &
  \left\{
  \begin{array}{ll}
  {\rm Black \ Hole:} &
  \alpha = \varphi = \gamma = \sigma = 3/2\, , \\[.4cm]
  {\rm Small \ Black \ Ring:} &
  \alpha = \varphi = \gamma = \sigma = 3/2\, , \\[.4cm]
  \beta = 1/\delta = -1/2\, .
  \end{array}
  \right.
  \end{array}
\end{equation}

Incidentally, let us notice that these critical exponents at
extremality precisely coincide with those found
in~\cite{KabuKerr,kabuscaling}, where four dimensional black holes
were considered. In that case, the two proposed phases were the inner
and outer horizons, the black hole solution being thought then as a
``coexistence curve''. In the five dimensional case we are
considering, our candidates for the two different phases are instead
the black ring and the black hole.

\subsection{Scaling Laws and Correlation Length in the BH/BR System}

With these critical exponents at our disposal, let us check now if the
scaling laws are obeyed. First of all, note that a nontrivial result
is the equality of the critical exponents $\alpha, \varphi,
\gamma$ and $\sigma$ at both phases of each transition point
considered (something which, in principle, need not be so). As for
the scaling relations involving them and the other critical
exponents defined so far, we quote here their expressions. They
are given by~\cite{ll}:
\begin{equation}
  \left\{
  \begin{array}{rcl}
    \alpha + 2\beta + \gamma & = & 2\, , \\[.2cm]
    \beta(\delta - 1) & = & \gamma\, , \\[.2cm]
    \varphi(\beta+\gamma) & = & \alpha\, .
  \end{array}
  \right.
\end{equation}
We can verify that these scalings are obeyed by the BH/BR system, both
at $x=x_{\rm min}$ and at $x=1$.

However, according to the general theory of critical phenomena,
further scaling laws involving additional critical exponents have to
be satisfied near a critical point~\cite{ll}. The remaining critical
exponents are those related to the behaviour of the two-point
correlation function and the correlation length. The two-point
correlation function $G(r)$ introduces a new critical exponent $\zeta$
defined by:
\begin{equation}
  G(r) \sim \frac{\exp(-r/\xi)}{r^{d-2-\zeta}}\, 
\end{equation}
($d$ is again the spatial dimension of the system). On the other
hand, the divergence of the correlation length $\xi$ at the critical
point is expressed in terms of the critical exponents $\mu$ and $\nu$
as:
\begin{equation}
  \xi \sim \epsilon_M^{-\nu} \sim \epsilon_J^{-\mu}\, .
\end{equation}
At a critical point these exponents obey the scaling laws given
by~\cite{ll}:
\begin{equation}
  \label{Scalings2Kind}
  \left\{
  \begin{array}{rcl}
  \nu(2-\zeta) & = & \gamma\, , \\[.2cm]
  2-\alpha & = & \nu d\, , \\[.2cm]
  \mu(\beta+\gamma) & = & \nu\, .
  \end{array}
  \right.
\end{equation}

To verify these relations we need to compute the correlation length
and the two-point correlator. However, there is no obvious way (not
even an obvious definition) to compute these quantities for a black
hole geometry. In any case, Eq.~(\ref{RuppAndCL}) provides us with an
explicit algorithm to compute the quantities $\nu d$ and $\mu d$ if
applicability of Ruppeiner theory is assumed. One obtains:
\begin{equation}
  \left\{
  \begin{array}{rl}
  {\rm At \ } x_{\rm min} {\rm :} & \ \
  \nu d = \mu d = 1\, , \\[.4cm]
  {\rm At \ } x=1 {\rm :} & \ \
  \nu d = \mu d = 1/2\, .
  \end{array}
  \right.
\end{equation}
Relations~(\ref{Scalings2Kind}) are obeyed at extremality for any
effective dimension $d$, but are not sa\-tisfied at $x_{\rm min}$. 

We therefore see that $x=1$ satisfies additonal scaling laws not
obeyed at $x_{\rm min}$. In this respect, we can conclude that the
behaviour of the BH/SBR system near extremality enjoys properties
which, formally, are analogous to those of a critical point of a
second order phase transition in which the SBR is always the
``disfavoured'' or metastable phase. As explained at the end of
Section~\ref{sec:DFAndCPh}, this statement has to be understood as
describing the properties of the system {\em only} along the stable
directions in configuration space.

On the other hand, at $x_{\rm min}$ we do not find an analogous
behaviour of the large and small ring phases. In spite of the
divergent behaviour of fluctuations, we see that the scalings related
to the critical exponents of the correlation length are not satisfied
there. This is in agreement with our expectations about the origin of
divergent fluctuations at $x_{\rm min}$ where, as we saw, what we have
is a change of stability. As discussed in Section~\ref{sec:DFAndCPh},
this phenomenon is very different from that of a critical point.


\section{Conclusions and Outlook}
\label{sec:Conclusions}

We have addressed the issues of stability, divergence of fluctuations
and critical phenomena in the BH/BR system.  Our results give a
consistent and unified picture of the properties of the phase diagram
of black holes and black rings. The two main tools that we have used
are the Poincar\'e method of stability and the geometry of the
thermodynamic state space. From a quite general point of view, we have
put special emphasis on the consequences of the non-extensivity of BH
thermodynamics, and we argued on the suitability of the methods used
here for the study of BHs. The main difference between the approach
followed in this paper and others in the BH literature is that ours is
based crucially on the study of the off-equilibrium function
$\widehat{S}(X^{\rho};\mu^i)$, and not on the properties of the
equilibrium entropy $S_{\rm eq}(\mu^i)$. The main advantage of the
Poincar\'e method is precisely that, just by plotting the right phase
diagrams of an {\em equilibrium} sequence, it can be used to infer
some relevant properties about $\widehat{S}(X^{\rho};\mu^i)$, even if
such a function is completely unknown. \newline

Let us he summarize the main results concerning the BH/BR
system. First, the small black ring has been shown to be locally
unstable. Concerning the large black ring, we found strong evidence
that this is a more stable configuration, i.e. one with less unstable
modes. It is nevertheless true that this does not imply its full
dynamical stability and, in fact, one expects such a configuration to
be unstable. The only place in the phase diagram of
Fig~\ref{fig:Entropies} where we see a change of stability is at
$x_{\rm min}$, where both BR branches meet.

We are able to evaluate thermodynamic fluctuations and compute the
second moments of correlations near $x_{\rm min}$ and $x=1$. In this
respect, we have discussed in detail the differences between turning
points and vertical asymptotes. We saw in Section~\ref{sec:DFAndCPh}
that, if one considers fluctuations only along the stable directions,
near a turning point the fluctuations of the conjugate variables are
dominated by the less stable mode $X^{\tilde{1}}$, i.e.
\begin{displaymath}
  \delta\widehat{\beta}_i \sim 
  (\partial_{\tilde{1}}\widehat{\beta}_i)_{\rm eq}\, 
  \Delta X^{\tilde{1}}\, ,
\end{displaymath}
due to the fact that $\Delta X^{\tilde{1}}$ is large. On the contrary,
near a vertical asymptote the fluctuations $\delta\widehat{\beta}_i$
are dominated by those in the usual thermodynamic variables, i.e.
\begin{displaymath}
  \delta\widehat{\beta}_i \sim 
  (\partial_j\widehat{\beta}_i)_{\rm eq}\, \Delta \mu^j\, ,
\end{displaymath}
due to the fact that $(\partial_j\widehat{\beta}_i)$ is divergent.
This is likely to be a more general phenomenon.

Finally, the computation of the various critical exponents allows to
formally interpret $x=1$ as a critical point along the stable
directions in configuration space, while at $x_{\rm min}$ the
divergence of fluctuations seems to be related only to a change of
stability. This fits with our expectations from the stability diagrams
in Fig.~\ref{fig:StabilityBHBR}, since at a turning point we only
expect a change in stability. Evidence of a critical point at a
vertical asymptote has also been found
in~\cite{KabuKerr,kabuscaling}. \newline

The extremal configurations at $x=1$ are naked singularities. However,
the general properties of a system at criticality concern its
behaviour near the critical point, and our configurations are
nonsingular there. A critical point indicates the endpoint of a
``coexistence curve'', along which a first order phase transition
between the two competing phases can take place.  It is true, however,
that in the BH/SBR system such a first order phase transition would
never be physically relevant in the dynamics of these spacetimes: not
only because the SBR is locally unstable, but also because this phase
always has lower entropy than the BH. Also, we do not know about any
phases beyond the ``critical point''. In any case, the fact that the
scaling relations are satisfied looks nontrivial. It would be
interesting to understand better the physical relevance of this
property.

On the other hand, the absence of any other points of change of
stability (in addition to the general arguments to rule out stability
considerations based on the sign of the specific heat) makes it very
unlikely to interpret the change of sign in the specific heat in the
BH branch as a phase transition. Moreover, at $x=1/2$ it is not clear
at all between which two phases there could be a transition.

Finally let us comment on the other special point in the phase diagram
of Fig.~\ref{fig:Entropies}: namely, $x=2\sqrt{2}/3$. Our results do
not indicate so far the existence of any special properties there,
neither in the BH nor in the LBR branch. Given this fact, we therefore
conclude that {\em if} both branches are stable around this point, the
only possibility we are left with is that of a first order phase
transition between both. We find no indications at all to interpret
$x=2\sqrt{2}/3$ as a critical point of a second order phase transition
between the BH and the LBR. \newline

From the point of view of the applicability of the techniques used in
this paper, it would be nice to understand in a more precise way the
relation between the off-equilibrium fluctuations $\{X^{\rho}\}$
considered here and the dynamical modes $\{\delta g_{\mu\nu}\}$ of
metric perturbations. Given the simplicity of the turning point
method, it would be very interesting to find an explicit and precise
connection, and see which dynamical modes these conjugacy diagrams are
actually probing. Even if, in general, one cannot expect to recover
full information on dynamical stability from the Poincar\'e method, it
seems clear that this method looks reliable at least to detect
instabilities. By this we mean that Poincar\'e instability (a negative
slope near a turning point) necessarily implies the existence of at
least one unstable dynamical mode (while the opposite is not
necessarily true).

In this respect, a natural question to ask is about the relation (if
any) of the Poincar\'e method used here and the Gubser-Mitra
conjecture~\cite{Gubser:2000mm}. The latter relates, for black brane
solutions, {\em standard} local thermodynamical instability to the
presence of dynamical instabilities in the same sense as discussed
here\footnote{That is, the existence of a classical instability under
Lorentzian time evolution.}. We have argued against the applicability
to BHs of the criterion of stability based on the sign of the various
specific heats. Note, however, that the GM conjecture does not include
compact horizons, just black branes with some {\em noncompact}
translational symmetry. This means that, along the noncompact
directions, one has extensivity of the mass per unit length and
additivity of the entropy per unit length. Therefore standard
thermodynamic criteria should apply in these cases, at least for
perturbations generating lumps along the uniform directions. Note that
these kind of GL instabilities are in perfect qualitative agreement
with the simple example that we discussed in the introduction of
Section~\ref{sec:DvsThStab} (a negative Hessian means that system is
unstable against redistribution of mass and tends to become
non-uniform). It would be extremely interesting to explore this issue
in more detail and see, in particular, if the appearance of this kind
of instabilities is related to bifurcations of equilibria.

Concerning the use of the thermodynamic metric and its relation to BH
physics, it would be desirable to have a precise definition and
understanding of the correlation length for a BH geometry. What we
have used here is the divergence of the thermodynamic curvature. This
is probably measuring the typical size of fluctuations in the BH
geometry as compared to the size of the BH itself. It would be nice to
work this out in an explicit~way.

The techniques we have been using are completely general, and obvious
extensions of this work include their application to other
gravitational phase transitions of interest. Work on the directions
proposed here is in progress.


\section*{Acknowledgments}

We would like to thank Ofer Aharony, Barak Kol, Niels Obers, Marika
Taylor and very specially Roberto Emparan for discussions on a
previous version of the manuscript; and Ofer Aharony, Daniel Cremades,
Gerard Iooss, Joseph Katz, Barak Kol and Robert Myers for discussions
and/or correspondence concerning the present version of this work.

E.~L.-T. wants to thank the Spinoza Institute for Theoretical Physics
of Utrecht~U. and the Institute of Theoretical Physics of K.~U.~Leuven
for hospitality during part of this work. 

The work of G.~A. has been supported by a Marie Curie Fellowship under
contract MEIF-CT-2003-502412, the ISF Israel Academy of Sciences and
Humanities, the GIF German-Israel Binational Science Found and the
European Network RTN HPRN-CT-2000-00122. The work of E.~L.-T. has been
supported by a Marie Curie Fellowship under contract
MEIF-CT-2003-502349, the Spanish Grant BFM2003-01090, The Israel-US
Binational Science Foundation, the ISF Centers of Excellence Program
and Minerva.


\appendix
\section{Fluctuations Near Extremality}
\label{app:FluctuationsAtExtremality}

Let us sketch here the derivation of equation~(\ref{dbidbjVA}) for
fluctuations near extremaility. In ordinary systems with positive
specific heat, the probability distribution just reduces to an
ordinary Gaussian integration due to the fact that the eigenvalues of
the Hessian matrix are all negative. Here, however, we always have one
positive eigenvalue, which makes the normalization constant of the
probability distribution strictly infinite if one is to integrate
fluctuations over the whole real line. Therefore, in evaluating the
expressions~(\ref{dbidbjVA}), we need to impose a cutoff in phase
space, i.e.
\begin{equation}
  \label{FluctuationsWithCutoff}
  \left\{
  \begin{array}{rcl}
  \langle\delta\widehat{\beta}_i\, \delta\widehat{\beta}_j\rangle\, 
  (\Lambda_M,\Lambda_J) 
  & = & \displaystyle \frac{1}{N(\Lambda_M,\Lambda_J)}
  \int_{-\Lambda_M}^{\Lambda_M}{d(\Delta M)}
  \int_{-\Lambda_J}^{\Lambda_J}{d(\Delta J)}
  \ \delta\widehat{\beta}_i\delta\widehat{\beta}_j 
  \ \exp(\widehat{S})\, , \\[.4cm] 
  N(\Lambda_M,\Lambda_J) 
  & = & \displaystyle\int_{-\Lambda_M}^{\Lambda_M}{d(\Delta M)}
        \int_{-\Lambda_J}^{\Lambda_J}{d(\Delta J)}
  \ \exp(\widehat{S})\, , 
  \end{array}
  \right.
\end{equation}
where $\widehat{S}$ stands for the quadratic approximation to the
entropy near the vertical asymptote,
i.e. Eq.~(\ref{DeltaSExtremality}), and the variations
$\delta\widehat{\beta}_i$ are given by
Eq.~(\ref{DeltaBetaExtremality}).

Before proceeding, let us note that a cutoff is always necessary: if
one is to make sense of any truncated series expansion in
$\Delta\mu^i$ of the entropy or $\delta\widehat{\beta}_i$, then the
variations $\Delta\mu^i$ cannot be arbitrarily large. This is also the
case when one has convergent Gaussian integrals. However one typically
has highly peaked distributions, and thus any physical cutoff is
naturally larger than the width of the Gaussian. Therefore a good
approximation which yields a simple analytic result is given by
integrating over the whole real line, which is what we always do. In
our case, the variations in $\Delta M$ and $\Delta J$ have to be
smaller than a certain quantity $\Lambda_{M,J}$ such that: a)~they are
much smaller than the total mass and angular momentum, b)~they are
small enough to be considered independent (since for large values they
would be related the extremality bound $x<1$), and c)~the truncated
series expansions~(\ref{DeltaBetaExtremality})
and~(\ref{DeltaSExtremality}) remain good approximations. The latter
condition can be used to give a simple estimation of the cutoff near
extremality, namely:
\begin{equation}
  \label{CutoffEstimation}
  \Lambda_{M,J}\sim {\cal O}(\epsilon)\, ,
\end{equation}
for $\epsilon=1-x$. This is obtained by computing the third
derivatives of the equilibrium entropy, which behave as
$\sim\epsilon^{-5/2}$ near the vertical asymptote. These give a simple
estimation of the order of magnitude of the next contribution in the
expansion of both $\widehat{S}$ and $\delta\widehat{\beta}_i$. Making
them negligible w.r.t. the second order contribution (see
Eqs.~(\ref{HessianBH}) and (\ref{HessianBRExpansion}))
yields~(\ref{CutoffEstimation}).
	
In our case we always have one positive and one negative
eigenvalue. Now, depending on which quantity we want to compute, we
can always arrange the integrand in such a way that the contribution
arising from the positive (``divergent'') mode cancels out in the
final result for any choice of the cutoff $\Lambda_{M,J}$. As a matter
of fact, one can check that:
\begin{equation}
  \label{dbdbdwdw}
  \left\{
  \begin{array}{rcl}
  \langle\delta\widehat{\beta}\, \delta\widehat{\beta}\rangle\, 
  (\Lambda_M=\infty,\Lambda_J) & = & 
  -H_{MM} + \beta^2\, , \\[.4cm]
  \langle\delta\widehat{\omega}\, \delta\widehat{\omega}\rangle\, 
  (\Lambda_M,\Lambda_J=\infty) & = & 
  -H_{JJ} + \omega^2\, .
  \end{array} 
  \right.
\end{equation}
This results are exact and do not require any explicit form of the
coefficients in the expansion of the entropy, only that $H_{MM}<0$,
$H_{JJ}<0$ and $\det H<0$, as in our case. The final dependence in the
cutoff that we retain finite in each case (which would give a
divergence if taken to be infinite) exactly cancels in the numerator
and the denominator of the general
expression~(\ref{FluctuationsWithCutoff}). To get this result we are
approximating, in each case, the integral over the negative
(``convergent'') mode with a finite cutoff with the integral over the
whole real line. To justify such an approximation, let us note that
the width $\sigma_i$ of the Gaussian integrals is given by ($i=M,J$)
\begin{equation}
  \sigma_i \sim 1/\sqrt{-H_{ii}}
  \sim {\cal O}(\epsilon^{3/4}) 
\end{equation}
near extremality (see Eqs.~(\ref{HessianBH}) and
(\ref{HessianBRExpansion})). This is always smaller than the
cutoff~(\ref{CutoffEstimation}).

Finally, the easiest way to find a simple approximation for the mixed
correlations 
$\langle\delta\widehat{\beta}\, \delta\widehat{\omega}\rangle$ 
is the following. One can check that:
\begin{equation}
  \langle(\beta+\delta\widehat{\beta})\, \delta\widehat{\omega}\rangle\, 
  (\Lambda_M=\infty,\Lambda_J) = 
  -H_{MJ}\, ,
\end{equation}
while, on the other hand:
\begin{equation}
  \langle\delta\widehat{\omega}\rangle\, 
  (\Lambda_M,\Lambda_J=\infty) = -\omega\, .  
\end{equation}
Since $\langle(\beta+\delta\widehat{\beta})\,
\delta\widehat{\omega}\rangle =
\beta\langle\delta\widehat{\omega}\rangle +
\langle\delta\widehat{\beta}\, \delta\widehat{\omega}\rangle$,
one gets:
\begin{equation}
  \label{dbdw}
  \langle\delta\widehat{\beta}\, \delta\widehat{\omega}\rangle = 
  -H_{MJ} + \beta\omega\, .
\end{equation}
Equations (\ref{dbdbdwdw}) and (\ref{dbdw}) complete therefore the
result~(\ref{dbidbjVA}) anticipated in Section~\ref{sec:FCBAndTG}.

Of course, the calculation above can also be carried out by using the
symmetric expressions:
\begin{equation}
  \langle(\omega+\delta\widehat{\omega})\, \delta\widehat{\beta}\rangle\, 
  (\Lambda_M,\Lambda_J=\infty) = -H_{MJ}\, , \ \ \ \ 
  \langle\delta\widehat{\beta}\rangle\, 
  (\Lambda_M=\infty,\Lambda_J) = -\beta\, .  
\end{equation}
Note the shifts in the mean fluctuations,
$\langle\delta\widehat{\beta}\rangle=-\beta$ and
$\langle\delta\widehat{\omega}\rangle=-\omega$, which are due to the
nonvanishing linear terms in the expansion of the entropy.



\end{document}